\documentclass[10pt,a4paper,onecolumn]{amsart}
\usepackage{etex}
\renewcommand{\baselinestretch}{1.05} 
\usepackage[utf8]{inputenc}
\usepackage[left=2.9cm,right=2.9cm,top=2.9cm,bottom=2.9cm]{geometry}

\usepackage{mathtools}
\usepackage{amssymb}
\usepackage{amsfonts}
\usepackage{amsthm}
\usepackage{stmaryrd}
\usepackage{multicol}
\usepackage{color}
\usepackage{graphicx}
\usepackage{setspace}

\usepackage{subcaption}
\expandafter\def\csname ver@subfig.sty\endcsname{}

\usepackage{dblfloatfix}

\usepackage{relsize}
\usepackage{verbatim}
\usepackage{subfig}

\usepackage{caption}

\usepackage{booktabs}
\usepackage{multirow}

\usepackage[
colorlinks=true,
linkcolor=black, 
anchorcolor=black,
citecolor=blue, 
filecolor=magenta, 
menucolor=red, 
urlcolor=cyan,
plainpages=false,
pdfpagelabels,
hypertexnames=false,
linktocpage 
]{hyperref}

\usepackage{tensorNotation}

\makeatletter
  \newcommand{\srcsize}{\@setfontsize{\srcsize}{5pt}{5pt}}
\def\smalloverbrace#1{\mathop{\vbox{\m@th\ialign{##\crcr
  \noalign{\kern.5\fontdimen5\textfont2}%
  \tiny\downbracefill\crcr
  \noalign{\kern.7\fontdimen5\textfont2\nointerlineskip}%
  $\hfil\displaystyle{#1}\hfil$\crcr}}}\limits}
  \def\tinyoverbrace#1{\mathop{\vbox{\m@th\ialign{##\crcr
  \noalign{\kern.5\fontdimen5\textfont2}%
  \srcsize\downbracefill\crcr
  \noalign{\kern.7\fontdimen5\textfont2\nointerlineskip}%
  $\hfil\displaystyle{#1}\hfil$\crcr}}}\limits}
\makeatother

\title[An Extended VoF Method for Bubbles Rising in a Viscoelastic Liquid]{An Extended Volume of Fluid Method and its Application to Single Bubbles Rising in a Viscoelastic Liquid}

\author[Niethammer et al.]{Matthias Niethammer$^{1}$, G{\"u}nter Brenn$^{2}$, Holger Marschall$^{1}$ and Dieter Bothe$^{1}$}

\address{$^{1}$Mathematical Modeling and Analysis, Center of Smart Interfaces, Technische Universit{\"a}t Darmstadt, Germany \\$^{2}$Institute of Fluid Mechanics and Heat Transfer, Graz University of Technology, Austria}

\email{niethammer@mma.tu-darmstadt.de}


\begin{document}

\begin{abstract}
An extended volume of fluid method is developed for two-phase direct numerical simulations of systems with one viscoelastic and one Newtonian phase. A complete set of governing equations is derived by conditional volume-averaging of the local instantaneous bulk equations and interface jump conditions. The homogeneous mixture model is applied for the closure of the volume-averaged equations. An additional interfacial stress term arises in this volume-averaged formulation which requires special treatment in the finite-volume discretization on a general unstructured mesh. A novel numerical scheme is proposed for the second-order accurate finite-volume discretization of the interface stress term. We demonstrate that this scheme allows for a consistent treatment of the interface stress and the surface tension force in the pressure equation of the segregated solution approach. Because of the high Weissenberg number problem, an appropriate stabilization approach is applied to the constitutive equation of the viscoelastic phase to increase the robustness of the method at higher fluid elasticity. Direct numerical simulations of the transient motion of a bubble rising in a quiescent viscoelastic fluid are performed for the purpose of experimental code validation. The well-known jump discontinuity in the terminal bubble rise velocity when the bubble volume exceeds a critical value is captured by the method. The formulation of the interfacial stress together with the novel scheme for its discretization is found crucial for the quantitatively correct prediction of the jump discontinuity in the terminal bubble rise velocity.
\end{abstract}

\maketitle



\section{Introduction}
\label{Introduction}
The hydrodynamics of single rising bubbles in a quiescent viscoelastic fluid matrix have been subject of numerous studies since the pioneering work of Astarita and Apuzzo \cite{Astarita1965}, who first reported that the bubbles may exhibit a jump discontinuity in the steady rise velocity when their volume exceeds a critical value. The critical bubble volume is characterized by a sharp increase in the steady rise velocity. This jump phenomenon can change the bubble rise velocity by about an order of magnitude, which makes it a subject of considerable importance in many process engineering applications, such as bubble columns. Astarita and Apuzzo related the jump in the bubble rise velocity to a change of shape of the rising bubble, i.e.\ they observed the formation of a ``teardrop'' shape which may cause the transition to a different flow regime. Hassager \cite{Hassager1979} observed a change of the flow field around the bubble, in particular the appearance of a downward flow in the central wake of the bubble, which is known in literature as the so-called negative wake effect. Later, flow measurements have substantiated the appearance of the negative wake \cite{Funfschilling2001}. The issue on the theoretical explanation for these three characteristic phenomena of rising bubbles in viscoelastic liquids led to a controversial debate over the past decades (for a review we refer to \cite{Chhabra2006, Zenit2018}).
Indeed, until now no complete theory has been proposed that explains the velocity jump discontinuity, and its derivation is not meant to be the goal of the present work also. Rather, the present work aims at the development of a numerical method, capable of predicting the velocity jump phenomenon and the flow field around the rising bubble in three dimensions. From that, we expect a detailed picture of the local state of stress around both subcritical and supercritical bubbles, which is hardly accessible in experiments. The numerical results for the steady bubble rise velocity are quantitatively compared with the well-characterized reference fluid and experimental data by Pilz and Brenn \cite{Pilz2007}.

A literature survey of previous numerical studies on a bubble rising in a quiescent viscoelastic fluid is shown in table \ref{tab:litsurvey}. The majority of studies published on this topic addresses the qualitative investigation of the bubble shape. Until now, finite element methods have been predominantly used in combination with both diffuse interface models \cite{Yue2006} and sharp interface models \cite{Pillapakkam2001, Pillapakkam2007}. Pillapakkam et al.\ \cite{Pillapakkam2007} reported the first numerical results for the velocity jump discontinuity and the negative wake effect. They presented both transient and steady state results of bubbles rising in a hypothetical polymer solution, modelled by the Oldroyd-B constitutive equations. Their simulations show a sharp increase in the steady rise velocity when the bubble exceeds a certain critical volume. Moreover, their results indicate that the bubble shape might become asymmetric, in particular for supercritical bubbles, which substantiates our intention to use a three-dimensional numerical approach. Recently, Fraggedakis et al.\ \cite{Fraggedakis2016} reported numerical results for the velocity jump discontinuity, using an axisymmetric arbitrary Lagrangian-Eulerian mixed finite element method. They approximated the reference fluid in \cite{Pilz2007} with an exponential Phan-Thien Tanner (EPTT) constitutive equation, obtaining good agreement with the experimental data in \cite{Pilz2007} for the subcritical rise velocities and bubble shapes. However, their numerical method suffers from stability issues when the bubbles exceeds the critical volume. To the best of our knowledge, numerical results for the velocity jump discontinuity have not previously been reported using the volume of fluid (VoF) method. Due to their inherent mass conservation property and robustness in handling strong morphological changes, the VoF method seems to be an appropriate choice for the simulation of rising bubbles in a viscoelastic fluid.
\begin{table}[h!]
\tiny
\begin{center}
\begin{tabular}{@{}llllll@{}}
\toprule
\multicolumn{3}{c}{} & \multicolumn{3}{c}{Subject of Investigation}\\ \cmidrule(l){4-6}
Reference & Numerical Method \ & Fluid Rheology \ & \ 
\begin{tabular}{@{}l@{}} 
Bubble\\
Shape
\end{tabular}
& \
\begin{tabular}{@{}l@{}} 
Negative\\
Wake
\end{tabular}
& \
\begin{tabular}{@{}l@{}} 
Jump\\
Discontinuity
\end{tabular}
\\ [-1ex]
\midrule
\cite{Noh1993} &
\begin{tabular}{@{}l@{}} 
axisymmetric \\ 
finite difference \\
method
\end{tabular}
& FENE-CR \ & \ 
x & \ & \ \\ [-1ex]
\midrule
\cite{Wagner2000} &
\begin{tabular}{@{}l@{}} 
two-dimensional \\ 
lattice Boltzmann \\
method
\end{tabular}
\ & Oldroyd-B \ & \ 
x & \ & \ \\ [-1ex]
\midrule
\cite{Pillapakkam2001} & 
\begin{tabular}{@{}l@{}} 
three-dimensional \\ 
level-set finite \\
element method
\end{tabular}
\ & Oldroyd-B \ & \ 
x & \ & \ \\ [-1ex]
\midrule
\cite{Sussman2005} & 
\begin{tabular}{@{}l@{}} 
axisymmetric \\ 
coupled level set\\
VoF method
\end{tabular} 
\ & Oldroyd-B \ & \ 
x & \ & \ \\ [-1ex]
\midrule
\cite{Yue2006} & 
\begin{tabular}{@{}l@{}} 
three-dimensional \\ 
phase-field finite \\
element method
\end{tabular} 
\ & Oldroyd-B \ & \ 
x & \ & \ \\ [-1ex]
\midrule
\cite{Malaga2007} & 
\begin{tabular}{@{}l@{}} 
axisymmetric \\ 
boundary integral \\
method
\end{tabular} 
\ & FENE-CR \ & \ 
x & \ & \ \\ [-1ex]
\midrule
\cite{Pillapakkam2007} & 
\begin{tabular}{@{}l@{}} 
three-dimensional \\ 
level-set finite \\
element method
\end{tabular} 
\ & Oldroyd-B \ & \ 
x & x & x \\
\midrule
\cite{Fraggedakis2016} & 
\begin{tabular}{@{}l@{}} 
axisymmetric \\ 
arbitrary Lagrangian-\\
Eulerian finite \\
element method
\end{tabular} 
\ & EPTT \ & \ 
x & x & (x)* \\
\bottomrule
\multicolumn{6}{l}{*method is unstable for supercritical bubbles} \\
\end{tabular}
\end{center}
\caption{\small Literature overview for numerical studies of rising bubbles in a quiescent viscoelastic fluid.}
\label{tab:litsurvey}
\end{table}

The VoF method \cite{Hirt1981} is an interface capturing method which is widely used for sharp interface direct numerical simulations (DNS). Besides the VoF method, several other popular interface capturing methods can be found in literature, such as the level-set method or the marker and cell method. In the VoF method, the sharp interface is captured by a phase indicator function. VoF methods can be classified depending on the advection scheme that is used to propagate the sharp interface. A common technique in literature is the geometrical advection based on an interface reconstruction technique. However, it is well-known that geometrical advection schemes are very difficult to implement in three dimensions on general unstructured grids \cite{Jofre2014, Maric2018}. On general unstructured grids, the most common practice is to advect the volumetric phase fraction algebraically with numerical schemes by solving a scalar transport equation. In this work, we follow the approach of an \textit{algebraic VoF}. This avoids the issues related to the complexity of the geometrical interface reconstruction and advection on unstructured grids, but it poses a huge challenge on the numerical schemes that need to keep the interface sharp and free of oscillations. Over the years, several numerical schemes for the algebraic advection of the sharp volume fraction field have been proposed from which we have to select a reasonable combination for the case of multi-dimensional flows on general unstructured grids. In section \ref{subsec:intcapt} we give a detailed discussion on the choice of suitable numerical schemes for the present study. We shall note that this choice is, even though being motivated by several physical and numerical arguments, not unique and different numerical schemes may work out as well.

The objectives of this work are to investigate the capability of the algebraic VoF method for the challenging issue of a rising bubble in a quiescent viscoelastic fluid matrix. First, we propose an extended VoF method for the two-phase system consisting of one viscoelastic and one Newtonian phase. Secondly, the model is validated by a comparison between simulation results and experimental data by \cite{Pilz2007}. We study whether the numerical method is capable of qualitatively predicting the characteristic phenomena mentioned above. Moreover, we provide a quantitative comparison between the measured steady state rise velocities and the simulation results. Finally, we present transient results for the start-up flow when a spherical bubble is initially at rest and starts to rise until a steady state is reached. For the implementation we use the OpenFOAM \cite{Weller1998} package, which provides an algebraic VoF solver (interFoam) for inelastic fluids on general unstructured meshes and, being written in C++, it offers the capability of a highly flexible and modular code design for our developments.

\section{Continuum Modeling}
\label{sec:contimod}
Let us consider a domain $\CCV(t) \subset \mathbb{R}^3$, and let $\Sigma(t) \subset \mathbb{R}^3$ be a two-dimensional smooth surface, cutting $\CCV$ into two subdomains (bulk phases), $\CCV^{\phaseOne}(t) \cup \CCV^{\phaseTwo}(t) \cup \Sigma_\CCV(t)$, with $\Sigma_\CCV := \Sigma \cap \CCV$. Let us further denote the interface unit normal pointing towards the phase $\CCV^{\phaseTwo}$ as $\vec{n}_{\Sigma}$.
\subsection{Bulk}
Assuming incompressible, isothermal and laminar flow, the local conservation equations for mass and momentum in the bulk read
\begin{align}
\label{contieq00}
\nabla \dprod \velocity &= 0 \quad \textnormal{in} \ \CCV \setminus \Sigma, \\
\label{monentumeq00}
\rho {\partial_t \velocity} + \rho \left(\velocity \dprod \nabla\right) \velocity &= \nabla \dprod \ST + \rho \vec{b} \quad \textnormal{in} \ \CCV \setminus \Sigma,
\end{align}
where $\velocity$ is the velocity, $\rho$ is the density and the term $\rho \vec{b}$ represents body forces with the mass specific density $\vec{b}$. The term $\nabla \dprod \ST$ represents contributions from contact forces, where $\ST$ is the Cauchy stress tensor of second rank. We split $\ST$ into an isotropic part, a viscous solvent component and an elastic polymer contribution
\begin{align}
\label{stressdef00}
\ST := -p \IT + \stressS + \stressP,
\end{align}
where $p$ is the pressure. Suppose that the viscous stress obeys Newton's law, i.e.\
\begin{align}
\label{stressdef01}
\stressS = 2 \svisc \DT,
\end{align}
where $\svisc$ is the solvent viscosity and $\DT = \frac{1}{2}\left(\grad{\velocity} + \trans{\grad{\velocity}}\right)$. For a multimode approximation of the relaxation spectrum, the elastic polymer contribution equals the combination of all $K$ modal polymer stress contributions
\begin{align}
\label{stressdef02a}
\stressP = \sum_{k = 0}^{K} \stressPk.
\end{align}
For a macroscopic fluid element, the modal polymer stress is governed by a statistical distribution of its internal microstructure, viz.\ the conformation of the macro-molecules; cf.\ \cite{Bird1977, Larson1988, Keunings2000}. Therefore, the modal polymer stress is directly related to a
conformation tensor $\Ck$, which may be regarded as the average distribution of all microscopic configurations of the polymer macromolecules in the macroscopic fluid element. In particular, the conformation tensor is usually related to the second moment of a distribution function of all molecular configurations \cite{Larson1988} which implies that, by definition, it is a symmetric and positive definite (spd) tensor of second rank. Indeed, the spd property of the conformation tensor is a universal requirement for the derivation of a class of numerical stabilization methods, applied in the present work, as will be discussed in section \ref{sec:stab}. In this work, we assume further that the modal polymer stress $\stressPk$ is a linear function of the symmetric and positive definite polymer conformation tensor $\Ck$, such that
\begin{align}
\label{stressdef02}
\stressPk = \frac{\pkvisc}{\tau_k (1-\zeta_{k})} \left( h_{0,k} \IT + h_{1,k} \Ck\right),
\end{align}
where $\pkvisc$ is the polymer viscosity, $\tau_{k}$ is the relaxation time and $\zeta_k$ is a material parameter (sometimes called ``slip parameter"). Note that (\ref{stressdef02}) may be derived from a closed form of the Kramers' expression for the polymer stress tensor \cite{Bird1987b, Larson1988}. The coefficients $h_{j,k}$ depend on the molecular model and the closure approximation. For most rheological models the scalar coefficients $h_{j,k}$ are either functions of the first invariant of $\Ck$, i.e.\ $h_{j,k} = \hat{h}_{j,k}(\tr{\Ck})$, or constants (cf.\ table \ref{tab:constitutiveEx}).
To obtain a complete system, additional constitutive equations for $\Ck$ are required. In this work, we restrict our considerations to partial differential conformation tensor constitutive equations of the form
\begin{align}
\label{constCEq00}
\partial_t \Ck + \left( \velocity \dprod \nabla \right)  \Ck  - \LT \dprod \Ck - \Ck \dprod \trans{\LT}
=
\frac{1}{\tau_{k}} \cvec{P}\left(\Ck \right) \quad \textnormal{in} \ \CCV \setminus \Sigma,
\end{align}
where $\LT := \trans{\grad \velocity} - {\zeta}_k \DT$ and the material parameter ${\zeta}_k \in [0, 2]$, characterizing the degree of non-affine response of the polymer chains to an imposed deformation. For ${\zeta}_k = 0$ the motion becomes affine in which case the left-hand side of (\ref{constCEq00}) reduces to the upper-convected derivative, first proposed by Oldroyd \cite{Oldroyd1950}. Let us note that (\ref{constCEq00}) is generic with respect to the model-specific fluid relaxation. The relaxation is introduced by a specification of the tensor function $\cvec{P} \left(\Ck \right)$. We impose the restriction that $\cvec{P} \left(\Ck \right)$ is a real analytic tensor function of $\Ck$, such that
\begin{equation}
\label{defisotropicf}
\OT \dprod \cvec{P} \left(\Ck \right) \dprod \trans{\OT} = \cvec{P} \left(\OT \dprod \Ck \dprod \trans{\OT} \right)
\end{equation}
for every orthogonal tensor $\OT$. In this case, $\cvec{P} \left(\Ck \right)$ is said to be an \textit{isotropic} tensor-valued function of the spd second-rank tensor $\Ck$. It has been shown \cite{RivlinEricksen1955, Rivlin1955} that every real analytic and isotropic tensor-valued function $\cvec{P} \left(\Ck \right)$ of one second-rank tensor $\Ck$ has a representation of the form
\begin{equation}
\label{Pfunc}
\cvec{P}\left(\Ck \right) = g_{0,k} \IT + g_{1,k} \Ck + g_{2,k} \Ck^2,
\end{equation}
where $g_{0,k}$, $g_{1,k}$ and $g_{2,k}$ are isotropic invariants (isotropic scalar functions) of $\Ck$ and hence can be expressed as functions of its three scalar principal invariants, i.e.\ $g_{i,k} = \hat{g}_{i,k}(I_{1,k}, I_{2,k}, I_{3,k}), \ i = 0,1,2$. The principal invariants $I_{1,k}, I_{2,k}, I_{3,k}$ are
\begin{equation}
I_{1,k} = \textnormal{tr} \: \Ck, \ I_{2,k} = \frac{1}{2} \left[ I_{1,k}^2 - \tr{\Ck^2} \right]\!, \ I_{3,k} = \det \Ck.
\end{equation}
We use only models for which $g_{2,k} = 0$, i.e.\ the quadratic term in (\ref{Pfunc}) vanishes. Table \ref{tab:constitutiveEx} summarizes the model-specific expressions for the constitutive equations, used in this work.
\begin{table}[h!]
\small
\begin{center}
\begin{tabular}{@{}lcccccc@{}}
\toprule
constitutive model & \ ${\zeta}_k$ \ & \ $g_{0,k}$ \ & \ $g_{1,k}$ \ & \ $g_{2,k}$ \ & \ $h_{0,k}$ \ & \ $h_{1,k}$ \ \\ 
\midrule
Maxwell/Oldroyd-B & $0$ & $1$ & $-1$ & $0$ & $-1$ & $1$ \\
LPTT & $\in \left[0, 2 \right]$ & $1 + \frac{\varepsilon_k}{1 - {\zeta}_k} \left( \tr \Ck - 3 \right) $ & $ - g_{0,k} $ & $0$ & $-1$ & $1$  \\  
EPTT & $\in \left[0, 2 \right]$ & $ \exp \left[  \frac{\varepsilon_k}{1 - {\zeta}_k} \left( \tr \Ck - 3 \right) \right] $ & $ - g_{0,k} $ & $ 0 $ & $-1$ & $1$ \\
\bottomrule
\end{tabular}
\end{center}
\caption{\small Model-dependent scalar-valued tensor functions for the generic constitutive equation (\ref{constCEq00}) and (\ref{Pfunc}).}
\label{tab:constitutiveEx}
\end{table}
\subsection{Interface}
\label{sec:interface}
The equations above are valid in the two bulk phases, $\CCV^{\phaseOne}$ and $\CCV^{\phaseTwo}$, respectively. At $\Sigma$, additional balance equations (jump conditions) are needed in order to obtain a complete system. Let us therefore define the jump $\jump{ \PhiT }$ of a quantity $\PhiT$ at $\Sigma$ as
\begin{align}
\label{defjump}
\jump{ \PhiT } (\vec{x}) = \lim\limits_{h \rightarrow 0^{+}} \left(\PhiT (\vec{x} + h\vec{n}_{\Sigma}) - \PhiT (\vec{x} - h\vec{n}_{\Sigma})\right), \quad \vec{x} \in \Sigma.
\end{align}
At the interface $\Sigma$ we impose the following conditions: first, there is no phase change, i.e.\ the mass flux across $\Sigma$ is zero. Secondly, there is no-slip of the bulk phases at $\Sigma$.
Then, the interfacial jump conditions are given by
\begin{align}
\label{jcontieqVV00}
\jump{ \velocity }  &= 0 \quad \textnormal{on} \ \Sigma, \\
\label{jmomentumeqVV00}
{\jump{\left(p \IT - \stressE\right)}} \dprod \vec{n}_{\Sigma} &= \sigma \kappa \vec{n}_{\Sigma} + \nabla_\Sigma \dprod \ST^\Sigma \quad \textnormal{on} \ \Sigma, \\
\label{snd00}
\velocity \dprod \vec{n}_\Sigma &= V_\Sigma \quad \textnormal{on} \ \Sigma,
\end{align}
where the local curvature is defined as $\kappa = \nabla_{\Sigma} \dprod (-\vec{n}_{\Sigma}) $, the extra stress reads $\stressE = \left(2\svisc \DT + \stressP\right)$ and $\ST^\Sigma$ is the interfacial stress tensor. In (\ref{jmomentumeqVV00}), $\nabla_\Sigma = \PT_{\Sigma} \dprod \nabla$ is the surface gradient operator, where $\PT_{\Sigma} := \IT - \vec{n}_{\Sigma} \otimes \vec{n}_{\Sigma}$ is the projection onto the local tangent plane. (\ref{snd00}) is the kinematic condition for the advection of the interface in absence of phase change, with $V_\Sigma$ denoting the speed of normal displacement of $\Sigma$.
For the interfacial stress tensor $\ST^\Sigma$ a constitutive model is needed. In a similar way as for the bulk phase, $\ST^\Sigma$ can be decomposed into 
\begin{align}
\label{split00}
\ST^\Sigma = \sigma \PT_{\Sigma} + \stressE^\Sigma,
\end{align}
where the first term on the r.h.s.\ contains the surface tension $\sigma$ which is assumed to be constant in the present work, i.e.\ we assume that there is no gradient driven flow on the interface. The term $\stressE^\Sigma$ represents the intrinsic interfacial extra stress that needs to be closed by a constitutive model for the rheology of the fluid interface. A non-zero $\stressE^\Sigma$ may arise from a possible adsorption of polymer molecules at the interface, developing a local microstructure along the interface, e.g.\ an interfacial monolayer. If the interfacial rheological stress in such an interfacial microstructure becomes sufficiently large compared to the bulk-fluid stresses acting across the interface, the intrinsic interfacial response needs to be considered by a rheological model for $\stressE^\Sigma$. In the present work, we assume that the microscale intrinsic interfacial stress is sufficiently small compared to the bulk-fluid stresses, such that it can be neglected in our macroscopic model, i.e.\ $\stressE^\Sigma = \vec{0}$. Since $\sigma$ is assumed to be constant, we obtain for the tangential contributions in (\ref{jmomentumeqVV00}) the condition
\begin{align}
\label{jmomentumeqVVt00}
- {\jump{ \PT_{\Sigma} \dprod \stressE \dprod \vec{n}_{\Sigma}}} &= \vec{0}.
\end{align}
Moreover, assuming that there is only one viscoelastic bulk fluid in phase $\phaseOne$, we have
\begin{align}
\label{jmomentumeqVVt00b}
- {\jump{ \PT_{\Sigma} \dprod \left(2\svisc \DT\right) \dprod \vec{n}_{\Sigma}}} + \PT_{\Sigma} \dprod \stressP_\phaseOne \dprod \vec{n}_{\Sigma} &= \vec{0}.
\end{align}
The component normal to the interface of (\ref{jmomentumeqVV00}) reads
\begin{align}
\label{jmomentumeqVVt01}
{\jump{p}} - {\jump{ \vec{n}_{\Sigma} \dprod  \left(2\svisc \DT \right) \dprod \vec{n}_{\Sigma}}} + \vec{n}_{\Sigma} \dprod \stressP_\phaseOne \dprod \vec{n}_{\Sigma} = \sigma \kappa.
\end{align}
Hence, the fluid interface is governed by the bulk-fluid stress contributions acting in the interface normal direction which are balanced by the surface tension force. For the shear stresses the slip condition (\ref{jmomentumeqVVt00b}) holds at the interface. In section \ref{sec:vofmodel}, we derive the volume-averaged (va)-VoF equations by applying a volume averaging procedure to the local instantaneous balance equations, given above. In the volume-averaged (va)-VoF equations, we use the momentum jump condition in the form
\begin{align}
\label{jmomentumeqVV03}
{\jump{\left(p \IT  - \left(2\svisc \DT \right)  \right)}} \dprod \vec{n}_{\Sigma} + \stressP_\phaseOne \dprod \vec{n}_{\Sigma} &= \sigma \kappa \vec{n}_{\Sigma} \quad \textnormal{on} \ \Sigma
\end{align}
as a closure relation.

\subsection{Stabilization}
\label{sec:stab} 
Because of the high Weissenberg number problem (HWNP) \cite{Joseph1985, Keunings1986}, a numerical stabilization is applied, in order to increase the robustness of the numerical method. Here, we use the unified mathematical and numerical stabilization framework for the general conformation tensor constitutive equations (\ref{constCEq00}) and (\ref{Pfunc}) as proposed in \cite{Niethammer2017}. The stabilization framework can be seen as a generalization of the logarithm conformation representation (LCR), first proposed by Fattal and Kupferman in \cite{Fattal2004}. In detailed validation studies of computational benchmarks, it has been found that the change-of-variable representations of the present stabilization framework improve the convergence at high Weissenberg numbers considerably \cite{Niethammer2017}. Below, we give a brief summary of the stabilization framework used in the present work. For a detailed description, we refer to \cite{Niethammer2017}. 

Let $\cvec{F}(\CT)$ be a real analytic tensor function of the spd second-rank tensor $\CT$, such that
\begin{equation}
\label{diagonalizef0}
\OT \dprod \cvec{F}(\CT) \dprod \trans{\OT}
=
\cvec{F}(\OT \dprod \CT \dprod \trans{\OT})
\end{equation}
for every orthogonal tensor ${\OT}$, i.e.\ let $\cvec{F}(\CT)$ be an \textit{isotropic} tensor-valued function of $\CT$. If $\CT$ is governed by the generic constitutive equation (\ref{constCEq00}), then $\cvec{F}(\CT)$ satisfies
\begin{equation}
\label{evolutionConf3}
\DDt {\cvec{F}(\CT)}
=
{2 \BT \dprod {\UpsilonT} \dprod \CT}
+
\OmegaT \dprod {\cvec{F}(\CT)} - {\cvec{F}(\CT)} \dprod \OmegaT
+
\frac{1}{\tau} \UpsilonT \dprod \cvec{P} \left(\CT \right),
\end{equation}
where we exploit the property that the spd conformation tensor is diagonalizable, i.e.\ $\CT = \QT \dprod \LambdaT \dprod \trans{\QT}$ with the diagonal tensor $\LambdaT$ containing the three real eigenvalues and the orthogonal tensor $\QT$ containing the corresponding set of eigenvectors. Then, $\cvec{F}(\CT) = \cvec{F}(\QT \dprod \LambdaT \dprod \trans{\QT}) = \QT \dprod \cvec{F}(\LambdaT) \dprod \trans{\QT}$.  Moreover, the deformation terms in the convective derivative can be decomposed into the first three terms on the r.h.s. of (\ref{evolutionConf3}), containing the tensors $\BT$ and $\OmegaT$ (see appendix \ref{asec:lddt} for the computation). Note that such a local decomposition, first proposed in \cite{Fattal2004}, is in general necessary to transform the convective derivative with some tensor function $\cvec{F}(\CT)$. Regarding the choice of the transformation function, we impose two fundamental requirements that $\cvec{F}(\CT)$ must satisfy in order to obtain both a numerically robust, but also a physically meaningful result: first, $\cvec{F}(\CT)$ must flatten the locally steep spatial conformation tensor profiles such that the stiffness of the advection problem is substantially reduced. Secondly, $\cvec{F}(\CT)$ must be chosen such that $\CT = \cvec{F}^{-1}\left(\cvec{F}(\CT)\right)$ preserves the positivity of the conformation tensor. A reasonable choice that satisfies both conditions is the $k$-th root of $\CT$, and the logarithm of $\CT$ to base $a$. Table \ref{tab:JJ}
\begin{table}[htbp]
\renewcommand{\arraystretch}{1.5}
\begin{center}
\linespread{1.2}
\begin{tabular}{rlll}
\toprule
\ & \cvec{F}(\CT) & \ $\UpsilonT$ \ \\
\midrule
conformation tensor \ & $\CT$ & \ $\IT$ \ \\
$k$-th root of $\CT$ \ & $\RT = \QT \dprod \LambdaT^{\frac{1}{k}} \dprod \trans{\QT}$ & \ $\frac{1}{k} \RT^{1-k}$ \ \\
logarithm of $\CT$ to base $a$ \ & $\ST = \QT \dprod \loga{(\LambdaT)} \dprod \trans{\QT}$ & \ $\frac{1}{\operatorname{ln}(a)} a^{-\ST}$ \ \\
\bottomrule
\end{tabular}
\end{center}
\renewcommand{\arraystretch}{1.0}
\caption{\small Function $\cvec{F}(\CT)$ and $\UpsilonT$ terms in (\ref{evolutionConf3}) for the conformation tensor representation, the $k$-th root of $\CT$ representation, and the logarithm of $\CT$ to base $a$ representation.}
\label{tab:JJ}
\end{table}
shows the corresponding $\UpsilonT$ terms for such functions.
The specific representations for $\RT$ and $\ST$ are then given by
\begin{align}
\label{evolutionConfRTb}
\DDt \RT
&=
\frac{2}{k} \BT \dprod \RT
+
\OmegaT \dprod {\RT} - {\RT} \dprod \OmegaT
+
\frac{1}{k \tau} \left(g_0 \RT^{1-k} + g_1 \RT + g_2 \RT^{1+k}\right),\\
\label{evolutionConfSTb}
\DDt \ST
&=
\frac{2}{\operatorname{ln}(a)} \BT
+
\OmegaT \dprod {\ST} - {\ST} \dprod \OmegaT
+
\frac{1}{\operatorname{ln}(a) \tau} \left(g_0 a^{-\ST} + g_1 \IT + g_2 a^{\ST}\right).
\end{align}
\section{VoF model}
\label{sec:vofmodel}
Within the finite volume framework, the method of volume-averaging \cite{Anderson1967, Slattery1967, Whitaker1967} is a well-known technique for the rigorous derivation of multiphase balance equations that are valid on the discrete level throughout the entire computational domain. In multiphase flows, this approach is usually called \textit{conditional} volume-averaging, because a phase-indicator is introduced to distinguish between the bulk phases before volume-averaging is applied.
For Newtonian two-phase flows, the concept of conditional volume-averaging has been successfully applied in the past for the derivation of Newtonian \textit{va-VoF} formulations \cite{Ubbink1997, Worner2001, Marschall2011}. In previous works, the Newtonian law was used as a closure model for the extra stress tensor in the momentum equation. Indeed, a similar closure is applied here for the viscous solvent part of the stress tensor of viscoelastic fluids. However, the elastic polymer stress tensor is modeled by partial differential constitutive equations which need to be written in a volume-averaged form, too. In this section, we propose a va-VoF formulation that contains additional contributions for the solvent and polymer stress of a viscoelastic phase.
\subsection{Volume-averaged VoF equations}
The volume-averaged VoF equations are obtained by averaging the local instantaneous bulk equations from section \ref{sec:contimod} over the volume $\Omega$. In order to distinguish between the two bulk phases $\phaseOne$ and $\phaseTwo$ in the domain $\Omega$, we introduce the phase-indicator function
\begin{align}
\chi_{i} (\vec{x}, t) := 
\left\lbrace
{
\begingroup
\renewcommand*{\arraystretch}{1.1}
\begin{array}{ll}
1 \ & \textnormal{if} \ \vec{x} \in \CCV^{i} \ \textnormal{at time t},
\\
0 \ & \textnormal{otherwise,}
\end{array}
\endgroup
}
\right.
\end{align}
where $i = \phaseOne, \phaseTwo$. The phase-indicator function describes the distribution of the bulk phases in $\Omega$. On the interface, $\chi_{i} (\vec{x}, t)$ is not defined although its left- and right-sided limits exist there. 
In the conditional volume-averaging approach, the local instantaneous bulk equations are multiplied with the phase-indicator function before volume averaging is applied. Detailed descriptions of the conditional volume-averaging technique can be found elsewhere \cite{Slattery1999, Whitaker1999, Drew1999, Jakobsen2008}. Here, we briefly introduce the definitions and operators of the conditional volume-averaging approach that are used in the derivation of the viscoelastic va-VoF equations of the present work. 
Let $\PhiT$ be some scalar or tensor quantity such that $\PhiT(t,\cdot)$ is continuous in $\CCV(t)$ and $\PhiT(\cdot, \vec{x})$ is continuously differentiable in $\CCV^\phaseOne \cup \CCV^{\phaseTwo}$. The \textit{volume average} of $\PhiT$ over $\CCV$ is defined as
\begin{align}
\label{defVA00}
\overline{\PhiT} := \frac{1}{\vert \Omega \vert} \int_{\Omega} \PhiT(\vec{x}, t) \dV
\end{align}
and the intrinsic phase average or \textit{phasic average} over $\CCV^{i}$ is defined as
\begin{align}
\label{defPA01}
\overline{\PhiT}^{i} := \frac{1}{\vert \CCV^{i} \vert} \int_{\CCV^{i}} \PhiT(\vec{x}, t) \dV.
\end{align}
Taking the volume average of the phase indicator function, we obtain the definition of the \textit{volumetric phase fraction} $\alpha_i$
\begin{align}
\label{defVF00}
\alpha_i := \overline{\chi_i} = \frac{1}{\vert \CCV \vert} \int_{\CCV} \chi_i(\vec{x}, t) \dV
= \frac{1}{\vert \CCV \vert} \int_{\CCV^{i}} 1 \dV = \frac{\vert \CCV^{i} \vert}{\vert \Omega \vert}.
\end{align}
The volumetric phase fraction is constrained by $\alpha_\phaseOne + \alpha_\phaseTwo = 1$ and $0 \leq \alpha_i \leq 1$.
With (\ref{defPA01}) and (\ref{defVF00}), we obtain for the conditional volume average of $\PhiT$ the relation
\begin{align}
\label{defCVA00}
\overline{\chi_i \PhiT} = \frac{\vert \CCV^{i} \vert}{\vert \CCV\vert} \frac{1}{\vert \CCV^{i} \vert} \int_{\CCV^{i}} \PhiT(\vec{x}, t) \dV = \alpha_i \overline{\PhiT}^{i}.
\end{align}
Similarly, we define the surface average of $\PhiT$ as
\begin{align}
\label{defSA00}
\smalloverbrace{\PhiT} = \frac{1}{\vert \Omega\vert} \int_{\Sigma_{\CCV}} \PhiT(\vec{x}, t) \dA,
\end{align}
where $\Sigma_{\CCV} := \Sigma \cap \CCV$ is the surface area inside $\CCV$.
Finally, we need the volume-averaged differentiation rules which address the exchange of time and space derivatives with the averaging operator \cite{Slattery1999, Whitaker1999, Drew1999, Jakobsen2008}:
\begin{align}
\label{defVAddt00}
\overline{\chi_i \partial_t \PhiT} 
&= \partial_t \left(\alpha_i \overline{\PhiT}^{i}\right) - \smalloverbrace{\PhiT_i \vec{u}^{\Sigma} \dprod \vec{n}_{i}}, \\
\label{defVAgrad00}
\overline{\chi_i \nabla \PhiT} 
&= \nabla \left( \alpha_i \overline{\PhiT}^{i} \right) + \smalloverbrace{\PhiT_i \vec{n}_{i}}, \\
\label{defVAdiv00}
\overline{\chi_i \nabla \dprod \PhiT} 
&= \nabla \dprod \left( \alpha_i \overline{\PhiT}^{i} \right) + \smalloverbrace{\PhiT_i \dprod \vec{n}_{i}},
\end{align}
where $\vec{n}_{i}$ is the outwardly directed unit normal of phase $i$, $\vec{u}^{\Sigma}$ is the velocity of the interface and $\PhiT_i$ is the value of $\PhiT$ in phase $i$.
Equation (\ref{defVAddt00}) is known as the Leibnitz rule for volume averaging, while (\ref{defVAdiv00}) represents the two-phase Gauss divergence theorem for volume averaging.  The differentiation rules (\ref{defVAddt00}) to (\ref{defVAdiv00}) have been rigorously derived, e.g.,\ by Anderson and Jackson \cite{Anderson1967}, Slattery \cite{Slattery1967} and Whitaker \cite{Whitaker1967, Whitaker1973, Whitaker1985}.
We use the rules above to reformulate the conditional volume averaged equations in their most common form as a combination of volume and interface terms.
Starting from the conditional volume-averaged mass balance
\begin{align}
\label{vofraceq00}
\overline{\chi_{i} \partial_t \rho} + \overline{\chi_{i}  \nabla \dprod (\rho \velocity)} = 0,
\end{align}
we derive a transport equation for the volumetric phase fraction. For this purpose, we apply (\ref{defVAddt00}) and (\ref{defVAdiv00}) to (\ref{vofraceq00}) and assuming that $\rho$ is constant in $\CCV^i$, we get
\begin{align}
\label{vofraceq01}
{\rho}_i \left( \partial_t \alpha_i + \nabla \dprod (\alpha_i \overline{\velocity}^i) \right)  = \rho_i \smalloverbrace{(\velocity^\Sigma - \velocity_i) \dprod \vec{n}_i},
\end{align}
where the surface integral on the r.h.s.\ is the phase specific mass flux from $\CCV^i$ across the interface which necessarily has to vanish in the absence of phase change. Employing further the continuity equation $\nabla \dprod \overline{\velocity}^i = 0$ yields 
\begin{align}
\label{vofraceq02}
\partial_t \alpha_i + \overline{\velocity}^i \dprod \nabla  \alpha_i  = 0.
\end{align}
Note that due to $\alpha_\phaseOne + \alpha_\phaseTwo = 1$, we need to consider only one transport equation (\ref{vofraceq02}) for either $\alpha_\phaseOne$ or $\alpha_\phaseTwo$.
The conditional volume-averaged momentum equation reads
\begin{align}
\label{vofUeq00}
\overline{\chi_{i} \partial_t \rho \velocity} + \overline{\chi_{i}  \nabla \dprod (\rho \velocity \velocity)} = \overline{\chi_{i}  \nabla \dprod \ST} + \overline{\chi_{i} \rho \vec{b}}.
\end{align}
Applying the differentiation rules above, we obtain
\begin{align}
\label{vofUeq01}
\partial_t (\alpha_i {\rho}_i \overline{\velocity}^i) + \nabla \dprod (\alpha_i {\rho}_i \overline{\velocity}^i \overline{\velocity}^i) = {\nabla \dprod (\alpha_i \overline{\ST}^i}) + \alpha_i {\rho}_i \vec{b} + \rho_i \smalloverbrace{\velocity_i (\velocity^\Sigma - \velocity_i) \dprod \vec{n}_i} + \smalloverbrace{\ST_i \dprod \vec{n}_i},
\end{align}
where the first surface integral on the r.h.s.\ vanishes in the absence of phase change, resulting in
\begin{align}
\label{vofUeq02}
\partial_t (\alpha_i {\rho}_i \overline{\velocity}^i) + \nabla \dprod (\alpha_i {\rho}_i \overline{\velocity}^i \overline{\velocity}^i) = {\nabla \dprod (\alpha_i \overline{\ST}^i)} + \alpha_i {\rho}_i \vec{b} + \smalloverbrace{\ST_i \dprod \vec{n}_i}.
\end{align}
Note that for the derivation of the convective term in (\ref{vofUeq01}) and (\ref{vofUeq02}) we have assumed that the volume average of the product of two quantities shall equal the product of two averaged quantities, i.e.\ we demand that all scales are well-resolved in the DNS, such that we can neglect possible dispersive fluxes caused by fluctuating velocity components.
Moreover, the interfacial terms of both phases $\phaseOne$ and $\phaseTwo$ on the r.h.s.\ in (\ref{vofUeq02}) are related to each other by the averaged form of the interface jump condition (\ref{jmomentumeqVV03}), i.e.\
\begin{align}
\label{vofIJ00}
\smalloverbrace{\ST_\phaseOne \dprod \vec{n}_\phaseOne} + \smalloverbrace{\ST_\phaseTwo \dprod \vec{n}_\phaseTwo}
= \smalloverbrace{{\jump{\left(p \IT  - \left(2\svisc \DT \right)  \right)}} \dprod \vec{n}_{\Sigma}} + \smalloverbrace{\stressP_\phaseOne \dprod \vec{n}_{\Sigma}}
= \sigma \smalloverbrace{\kappa \vec{n}_{\Sigma}}.
\end{align}
\subsection{Closure}
The homogeneous mixture model is a widely used closure for volume averaged two-phase flow models in which it is assumed that the two phases may be treated as a single homogeneous continuum. 
The concept is based on the \textit{one-field} assumption for the balance equations. That is, the volume averaged balance equations for mass and momentum of the two phases are reformulated by \textit{one} set of balance equations, valid for both phases within the whole computational domain. This homogeneous mixture model is a good assumption if the following requirements are satisfied: First, both phases are assumed to move with the local center of mass velocity. Hence, the local volume averaged relative velocity $\velocity_r = \overline{\velocity}^\phaseTwo - \overline{\velocity}^\phaseOne $ between the two phases is assumed to be zero within the averaging volume $\CCV$. Secondly, all relevant scales of the flow and the interface topology are assumed to be resolved, i.e.\ the model requires a DNS.
We apply the homogeneous mixture model in this work, keeping in mind that we have to set up the simulations such that the interface topology is sufficiently well resolved. The mixture model equations are derived by taking the sum of the conditional volume averaged equations for both phases in the system. This results in
\begin{align}
\label{mixContieq00}
\nabla \dprod {\velocity}_m &= 0, \\
\label{vofraceq02b}
\partial_t \alpha_\phaseOne + {\velocity}_m \dprod \nabla  \alpha_\phaseOne  &= 0, \\
\label{mixUeq00}
\partial_t {\rho}_m {\velocity}_m 
+ ({\velocity}_m \dprod \nabla) {\rho}_m {\velocity}_m
&= {\nabla \dprod \left(\alpha_\phaseOne \overline{\ST}^\phaseOne \right)} + {\nabla \dprod \left(\alpha_\phaseTwo \overline{\ST}^\phaseTwo \right)} + {\rho}_m \vec{b} + \sigma \smalloverbrace{\kappa \vec{n}_{\Sigma}},
\end{align}
where we have used the following definitions for the mixture density $\rho_m$ and the center of mass velocity $\velocity_m$, respectively:
\begin{align}
\label{mixdensity00}
\rho_m = \alpha_\phaseOne \rho_\phaseOne + \alpha_\phaseTwo \rho_\phaseTwo,
\end{align}
\begin{align}
\label{mixvelocity00}
\velocity_m = \frac{\alpha_\phaseOne \rho_\phaseOne \overline{\velocity}^\phaseOne + \alpha_\phaseTwo \rho_\phaseTwo \overline{\velocity}^\phaseTwo}{\rho_m}.
\end{align}
For the derivation of (\ref{mixContieq00}) - (\ref{mixUeq00}), the mixture model assumption has been applied, stating that the local volume averaged relative velocity must vanish, i.e.\ $\velocity_r = 0 $ in $\CCV$.
In (\ref{mixUeq00}), the momentum jump condition (\ref{vofIJ00}) has been employed to express the sum of the interface stress terms. After the application of the mixture model, the system is still not complete. Additional closure assumptions are required to express the stress tensor terms and the surface tension force term on the r.h.s\ of (\ref{mixUeq00}). The next subsection proposes how to integrate the constitutive relationships for the stress into the mixture model equations.
\subsubsection{Constitutive equations for the stress tensor}
According to (\ref{stressdef00}), the divergence of the stress tensor in (\ref{mixUeq00}) may be split into three contributions, viz.
\begin{align}
\label{closestress00}
\sum_i \nabla \dprod \left(\alpha_i \overline{\ST}^i\right) = - \sum_i \nabla \left(\alpha_i \overline{p}^i\right) + \sum_i \nabla \dprod \left(\alpha_i \overline{\stressS}^i\right) + \nabla \dprod \left(\alpha_\phaseOne \overline{\stressP}^\phaseOne\right), \quad i = \phaseOne, \phaseTwo.
\end{align}
The first term on the r.h.s\ in (\ref{closestress00}) contains the volume averaged pressure. The homogeneous mixture model demands that there is \textit{one} pressure field which is shared by both phases in $\CCV$. Hence, it has become customary to denote the pressure as
\begin{align}
\label{closestress01}
p := \overline{p}^\phaseOne = \overline{p}^\phaseTwo,
\end{align}
from which it follows that
\begin{align}
\label{closestress02}
- \sum_i \nabla \left(\alpha_i \overline{p}^i\right) = -\nabla p.
\end{align}
The second term on the r.h.s\ in (\ref{closestress00}) contains the volume averaged purely viscous stress tensor. In accordance with (\ref{stressdef01}), we apply the constitutive equations of a Newtonian fluid to the viscous solvent to obtain
\begin{align}
\label{closestress03}
\sum_i \nabla \dprod \left(\alpha_i \overline{\stressS}^i\right) = 
\sum_i \nabla \dprod \left[{\svisc}_i \alpha_i \left(\grad{\overline{\velocity}^i} + \trans{\grad{\overline{\velocity}^i}}\right)\right], \quad i = \phaseOne, \phaseTwo.
\end{align}
In order to close (\ref{closestress03}), we search an expression in terms of the center of mass velocity (\ref{mixvelocity00}). Defining 
\begin{align}
\label{closestress04}
{\stressSm} = {\svisc}_m \left(\grad{{\velocity}_m} + \trans{\grad{{\velocity}_m}}\right),
\end{align}
where the mixture solvent viscosity is defined as
\begin{align}
\label{mixviscosity00}
{\svisc}_m := \alpha_\phaseOne {\svisc}_\phaseOne + \alpha_\phaseTwo {\svisc}_\phaseTwo
\end{align}
and comparing with (\ref{closestress03}), it turns out that for the mixture model the relationship between (\ref{closestress03}) and (\ref{closestress04}) is given by 
\begin{align}
\label{closestress05}
\nabla \dprod {\stressSm} =
\sum_i \nabla \dprod \left[ {\svisc}_i \alpha_i \left(\grad{\overline{\velocity}^i} + \trans{\grad{\overline{\velocity}^i}}\right)\right] .
\end{align}
Note that (\ref{closestress05}) is satisfied only if the phase relative velocity $\velocity_r$ is assumed to be zero. For the implementation into a numerical code, the divergence term $\nabla \dprod {\stressSm}$ is written in a more convenient form according to
\begin{align}
\nabla \dprod {\stressSm} &=
\nabla \dprod \left[{\svisc}_m \left(\grad{{\velocity}_m} + \trans{\grad{{\velocity}_m}}\right)\right]
\nonumber
\\
\label{closestress06b}
&=
\nabla \dprod \left({\svisc}_m \grad{{\velocity}_m}\right) + \grad{{\velocity}_m} \dprod \nabla {\svisc}_m,
\end{align}
where we used (\ref{mixContieq00}), that is the divergence of the barycentric velocity ${\velocity}_m$ is zero. The Laplacian term on the r.h.s.\ in (\ref{closestress06b}) is discretized implicitly in this work.
Finally, we consider the third term on the r.h.s\ in (\ref{closestress00}) containing the volume averaged polymer stress tensor. This term is present only in phase $\phaseOne$. Due to the stabilization approach for high Weissenberg numbers \cite{Niethammer2017}, the constitutive equations for the polymer stress are transformed to an equivalent set of equations with respect to the conformation tensor as (\ref{stressdef02}). Conditional volume-averaging of the algebraic relation (\ref{stressdef02}) yields
\begin{align}
\label{pstressva01}
\alpha_\phaseOne \overline{\stressP}^\phaseOne &= \alpha_\phaseOne \frac{{\pvisc}_\phaseOne}{\tau_\phaseOne (1 - \zeta_\phaseOne)}\left(\overline{\CT}^\phaseOne - \IT \right).
\end{align}
The conformation tensor is modeled by the partial differential constitutive equation in the general form (\ref{constCEq00}). Conditional volume-averaging of (\ref{constCEq00}) yields
\begin{align}
\label{constCEqva00}
\overline{\chi_{\phaseOne}\partial_t \CT} + \overline{\chi_{\phaseOne}\left( \velocity \dprod \nabla \right)  \CT}  - \overline{\chi_{\phaseOne}\LT \dprod \CT} - \overline{\chi_{\phaseOne}\CT \dprod \trans{\LT}}
=
\overline{{\frac{\chi_{\phaseOne}}{\tau_{\phaseOne}} \cvec{P}\left(\CT \right)}}.
\end{align}
Employing the differentiation rules (\ref{defVAddt00}) to (\ref{defVAdiv00}) and assuming a well-resolved DNS in which dispersion and fluctuation terms are negligibly small, we find
\begin{align}
\label{constCEqva01}
&\partial_t \left(\alpha_\phaseOne \overline{\CT}^\phaseOne \right) + \left( \overline{\velocity}^\phaseOne \dprod \nabla \right) \alpha_\phaseOne \overline{\CT}^\phaseOne  - \alpha_\phaseOne \overline{\LT}^\phaseOne \dprod \overline{\CT}^\phaseOne - \alpha_\phaseOne \overline{\CT}^\phaseOne \dprod {\trans{(\overline{\LT}^\phaseOne)}}
= 
{\frac{\alpha_\phaseOne}{\tau_\phaseOne} \cvec{P}\left(\overline{\CT}^\phaseOne \right)},
\end{align}
where we assume that
\begin{align}
\label{constCEqva01b}
\smalloverbrace{{\CT}_\phaseOne (\velocity^\Sigma - \velocity_\phaseOne) \dprod \vec{n}_\phaseOne}
= \vec{0},
\end{align}
since there is no phase change. The volume-averaged equations (\ref{pstressva01}) and (\ref{constCEqva01}) constitute the basis for the closure of the mixture model. In order to derive a closure, it is significant that we are facing a slightly different problem than above. We know a priori that the polymer stress exists only in phase $\phaseOne$ and is zero in phase $\phaseTwo$. Moreover, as described in section \ref{sec:interface}, no intrinsic interfacial rheology is considered. Therefore, we consider whether the definition of a complete set of mixture quantities is the most adequate strategy for the closure of the present system. Introducing a mixture stress implies the assumption that it appears in both phases, at least during the derivation. The result can then be applied, assuming a zero stress tensor field in one of the two phases. However, since this is more general than we need, we try to avoid the issue of finding a proper closure for the problem when both phases are viscoelastic. Due to the complicated structure of the constitutive relationships in the general (model-independent) mathematical framework of this work, it is a rather challenging task to arrive at a complete set of appropriate definitions for all quantities in question. To obtain a complete system, an alternative approach for the specific situation, where we have one polymeric phase $\phaseOne$, appears to be more straightforward. Without the aim of finding a proper definition for all mixture quantities, we assume that there are two polymeric phases and define $\CT_m := \alpha_\phaseOne \overline{\CT}^\phaseOne + \alpha_\phaseTwo \overline{\CT}^\phaseTwo$. 
Summing up the constitutive equations for both phases, we find after some algebraic manipulations (see appendix \ref{asec:mcte})
\begin{align}
\nonumber
&\partial_t {\CT}_m + \left( \velocity_m \dprod \nabla \right){\CT}_m  - {\LT}_m \dprod {\CT}_m - {\CT}_m \dprod \trans{({\LT}_m)} \\
\label{constCEqva01d}
= \; &
{\frac{\alpha_\phaseOne}{\tau_\phaseOne} \cvec{P}\left(\overline{\CT}^\phaseOne \right)}
+
{\frac{\alpha_\phaseTwo}{\tau_\phaseTwo} \cvec{P}\left(\overline{\CT}^\phaseTwo \right)}
,
\end{align}
which is valid under the mixture model assumption $\velocity_r = \vec{0}$.
Inserting $\CT_{\phaseTwo} = \vec{0}$ into (\ref{constCEqva01d}) we find the constitutive equation for phase $\phaseOne$
\begin{align}
\label{constCEqva01e}
&\partial_t \left(\alpha_\phaseOne \overline{\CT}^\phaseOne \right) + \left( \velocity_m \dprod \nabla \right) \alpha_\phaseOne \overline{\CT}^\phaseOne  - \alpha_\phaseOne {\LT}_m \dprod \overline{\CT}^\phaseOne - \alpha_\phaseOne \overline{\CT}^\phaseOne \dprod \trans{({\LT}_m)}
=
{\frac{\alpha_\phaseOne}{\tau_\phaseOne} \cvec{P}\left(\overline{\CT}^\phaseOne \right)}
\end{align}
for the mixture model. Equation (\ref{constCEqva01e}) may be inserted together with the algebraic relation for the polymer stress (\ref{pstressva01}) into the stress divergence term of the momentum balance. Furthermore, by inserting the phase fraction transport equation (\ref{vofraceq02}) into (\ref{constCEqva01e}), we find
\begin{align}
\label{constCEqva02}
&\partial_t \overline{\CT}^\phaseOne + \left( \velocity_m \dprod \nabla \right)  \overline{\CT}^\phaseOne  - \alpha_\phaseOne {\LT}_m \dprod \overline{\CT}^\phaseOne - \alpha_\phaseOne \overline{\CT}^\phaseOne \dprod \trans{({\LT}_m)} =
{\frac{\alpha_\phaseOne}{\tau_\phaseOne} \cvec{P}\left(\overline{\CT}^\phaseOne \right)}.
\end{align}
With respect to (\ref{pstressva01}), the transport of the phasic average $\overline{\CT}^\phaseOne$ instead of $\alpha_\phaseOne \overline{\CT}^\phaseOne$ is equivalent. Solving the constitutive equation for the phasic average of the conformation tensor (\ref{constCEqva02}) would be a legitimate choice for a limited range of Weissenberg numbers. However, a change-of-variable representation similar to (\ref{evolutionConf3}) is significantly more robust over a wide range of Weissenberg numbers and, therefore, is being pursued. To set up the change-of-variable representation, the constitutive equation (\ref{constCEqva02}) must be transformed as shown in section \ref{sec:stab}. For the volume-averaged change-of-variable representation of (\ref{constCEqva02}) we finally arrive at
\begin{align}
\nonumber
&\partial_t {\cvec{F}(\overline{\CT}^\phaseOne)} + \left( \velocity_m \dprod \nabla \right)  {\cvec{F}(\overline{\CT}^\phaseOne)} \\
= \; &
\label{constCEqva03}
\alpha_\phaseOne \left(
{2 {\BT}_m \dprod \overline{\UpsilonT}^\phaseOne \dprod \overline{\CT}^\phaseOne }
+
{\OmegaT}_m \dprod {\cvec{F}(\overline{\CT}^\phaseOne)} - {\cvec{F}(\overline{\CT}^\phaseOne)} \dprod {\OmegaT}_m
+
\frac{1}{\tau_\phaseOne} \overline{\UpsilonT}^\phaseOne \dprod \cvec{P} (\overline{\CT}^\phaseOne ) \right).
\end{align}
The representation (\ref{constCEqva03}) is the volume-averaged counterpart for the mixture model of the generic change-of-variable representation (\ref{evolutionConf3}) for a single-phase system. It is valid only if phase $\phaseOne$ is viscoelastic, while phase $\phaseTwo$ is not. The application of a certain tensor function $\cvec{F}(\overline{\CT}^\phaseOne)$ can affect the numerical stability, as described in section \ref{sec:stab}. According to the benchmark results \cite{Niethammer2017}, a reasonable compromise between efficiency, stability and accuracy is the 4th-root tensor function, which is used in this work. An essential advantage of the root conformation tensor representation is that the back-transformation of the generic transport variable $\vec{G} := \cvec{F}(\overline{\CT}^\phaseOne)$ to the polymer stress tensor does not necessarily require a diagonalization of the conformation tensor, which is exploited here to reduce the computational costs.

Equations (\ref{pstressva01}), (\ref{constCEqva03}) and the transformations in Table \ref{tab:JJ} constitute the closure for the polymer stress tensor. Having a complete system, we consider one further aspect of the polymer stress divergence term of the momentum balance. Decomposing the polymer stress term with the product rule yields
\begin{align}
\label{closestress07}
\nabla \dprod \left(\alpha_\phaseOne \overline{\stressP}^\phaseOne\right) =  \alpha_\phaseOne \nabla \dprod \overline{\stressP}^\phaseOne + \overline{\stressP}^\phaseOne \dprod \nabla \alpha_\phaseOne.
\end{align}
This decomposition may be used to separate pure interface contributions in the second term on the r.h.s.\ of (\ref{closestress07}) from the remainder before some numerical discretization is applied. Inserting $\PhiT = 1$ into (\ref{defVAgrad00}), it is easy to see that 
\begin{align}
\label{closestress08}
\nabla \alpha_\phaseOne = \smalloverbrace{\vec{n}_{\Sigma}},
\end{align}
demonstrating the connection between the gradient of the phase fraction and the surface average of the interface normal vector.  Even though the l.h.s.\ and the r.h.s.\ of (\ref{closestress07}) are identical from a mathematical point of view, they are not after certain numerical differencing schemes are applied. In this respect, it is beneficial to employ the same differencing scheme for the gradient of the phase fraction in the interface term on the r.h.s.\ of (\ref{closestress07}) as for the surface tension force term. In this way, the directional consistency between the contributions acting in interface normal direction can be realized on the discrete level. The discretization practice of (\ref{closestress07}) is described in detail in section \ref{sec:nummeth}.

\subsubsection{Surface tension force}
In the following, we briefly describe the incorporation of the surface tension in (\ref{mixUeq00}). Using the relationship (\ref{closestress08}), the surface tension force in (\ref{mixUeq00}) can be reformulated as
\begin{align}
\label{stf00}
\vec{f}_\sigma = \sigma \smalloverbrace{\kappa \vec{n}_{\Sigma}} = \sigma \smalloverbrace{\kappa} \nabla \alpha_\phaseOne.
\end{align}
This formulation was proposed by Brackbill et al.\ \cite{Brackbill1992} in the widely used continuous surface force (CSF) model, which shall be employed here. The CSF model is based on the idea that
the surface tension force acting on the sharp interface may be replaced by a continuous volumetric force density acting locally within the interfacial transition region $0 < \alpha_\phaseOne < 1$. 
The curvature in (\ref{stf00}) is approximated as
\begin{align}
\label{stf01}
{\smalloverbrace{\kappa}} = {\smalloverbrace{ \nabla_\Sigma \dprod (- \n_\Sigma)}}
\approx -\nabla \dprod \left(\frac{\nabla \alpha_\phaseOne}{\vert \nabla \alpha_\phaseOne \vert}\right),
\end{align}
which is consistent with the original CSF model. 

\section{Numerical method}
\label{sec:nummeth}

\subsection{Finite volume discretization}
\subsubsection{Domain discretization}
\begin{figure}[htbp!]
\centering{
\includegraphics[width=189pt]{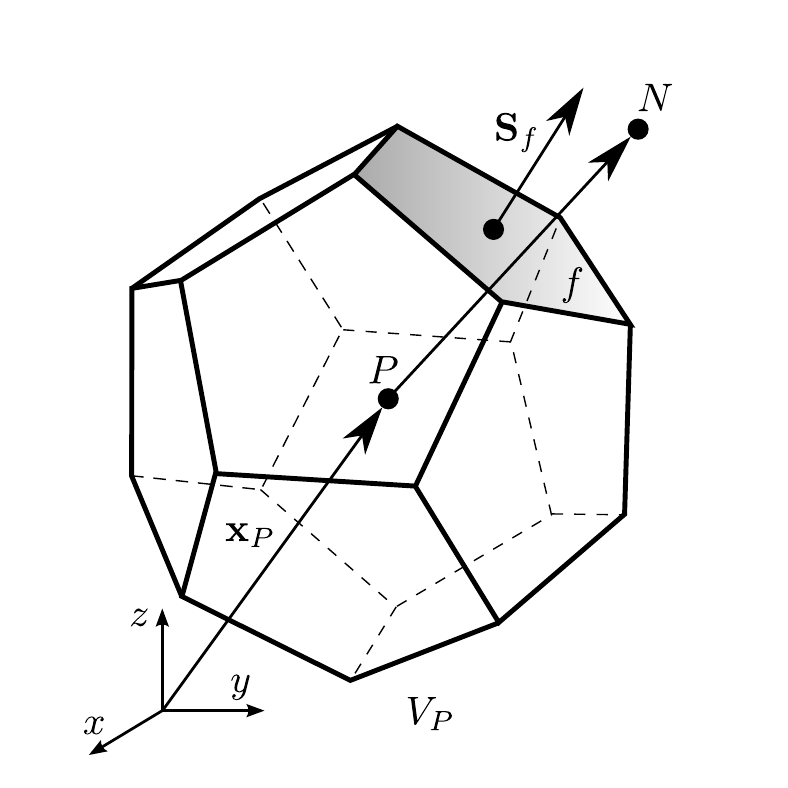}}
\caption{\small A generic control volume $V_P$ around the computational point $P$.}
\label{fig:polyCV}
\end{figure}

The space discretization is carried out by a subdivision of the computational domain into a set of convex control volumes (CVs), that do not overlap and completely fill the computational domain. Fig.\ \ref{fig:polyCV} 
shows a generic CV $V_P$ of general polyhedral shape around a computational point $P$ that is located in its centroid. 
$V_P$ is bounded by a set of flat faces $f$. $\vec{s}_f$ is defined for each face as the outward-pointing vector normal to $f$ with the magnitude of the surface area of $f$, i.e.\ $\vec{s}_f = \vec{n}_f\vert \vec{s}_f \vert$, where $\vec{n}_f$ is the face unit normal vector. Each face of $V_P$ is shared with one neighboring CV.

\subsubsection{Equation discretization}
In the discretization of the viscoelastic VoF equations proposed in section \ref{sec:vofmodel} on a general unstructured mesh, special care must be taken to the treatment of the polymer stress term at the fluid interface. To satisfy the interface balance in the volume-averaged framework, the discretization of the interface stress term and the surface tension force term should be carried out on the same stencil in the pressure equation of our segregated solution approach. Moreover, since we are using a collocated variable arrangement in the computational grid where all dependent variables are stored in the center of the CVs, the gradients of the volume-fraction in the interface terms require special cell-face interpolation techniques, as proposed by Rhie and Chow \cite{RhieChow1983}, to prevent unphysical checkerboard pressure and velocity fields. For the surface tension force at the cell-face, it is well-known that if the cell-face gradient along the $PN$ direction is discretized on a compact stencil in terms of the adjacent cell-centroid variables, such decoupling can be removed. The key aspect of the present section is to propose a similar discretization practice for the interface polymer stress term in the pressure equation of the segregated solution approach. The interface polymer stress is discretized at the cell-face of the CVs on a compact stencil to prevent decoupling and to obtain consistency with the discretization of the surface tension force at the cell-face.
In the following, we drop the index $m$ of the mixture quantities in the notation for the sake of better readability.
A second-order accurate approximation in space and time of the discretized momentum equation (\ref{mixUeq00}) for the CV of Fig.\ \ref{fig:polyCV} can be written as
\begin{align}
\nonumber
&\frac{ V_P}{\vartriangle \! t} \rho \left(\velocity^n - \velocity^o\right)
+ \sum_f {\rho}_f \phi_f^{\beta^*} {\velocity}_f^{\beta}
- \sum_f {{{\svisc}}_f}^{\beta^*} \vec{s}_f \dprod \left(\nabla {\velocity}\right)_f^{\beta}\\
= \; &
\label{momentumeqdiscr01}
\sum_f \left(\vec{s}_f {\velocity}_f^{\beta^*}\right) \dprod \sum_f \left(\vec{s}_f {\svisc}_f^{\beta^*}\right)
+ \alpha_\phaseOne \sum_f \vec{s}_f \dprod (\overline{\stressP}^\phaseOne)_f^{\beta^*}
+ {V_P} \vec{q}^{\beta^*},
\end{align}
where $\vec{q}$ represents some explicitly evaluated source terms that shall be discretized at a later stage,
\begin{align}
\label{momentumeqdiscr01source}
\vec{q} :=
-\nabla p 
+ {\rho} \vec{b} + \sigma \smalloverbrace{\kappa} \nabla \alpha_\phaseOne
 + \overline{\stressP}^\phaseOne \dprod \nabla \alpha_\phaseOne,
\end{align}
and $\phi_f = \vec{s}_f \dprod {\velocity}_f$ is the volumetric flux.
The quantity $\beta$ is the temporal off-centering coefficient in the interpolation between the old time steps $o$ and the new time steps $n$, i.e.\ for any scalar, vector or tensor quantity $\PhiT_f^{\beta}$ we have
\begin{equation}
\label{beta1}
\PhiT_f^{\beta} = \frac{1}{1+{\beta}} \PhiT_f^n + \frac{{\beta}}{1+{\beta}} \PhiT_f^o.
\end{equation}
In (\ref{beta1}), a value of ${\beta}=1$ equals a second-order Crank-Nicolson scheme, while ${\beta}=0$ results in a first-order Euler scheme. In this work, we use a small amount of off-centering, i.e.\ ${\beta}=0.95$, to increase the stability of the Crank-Nicolson scheme. Since we are using a segregated solution approach, some components of (\ref{momentumeqdiscr01}) cannot be discretized fully implicitly, such as the non-linear convection term and the source terms. We mark quantities that are lagged by one outer iteration at the current time level as $\PhiT_f^{n^*}$. For a lagged quantity, the temporal interpolation (\ref{beta1}) can be reformulated as
\begin{equation}
\label{beta2}
\PhiT_f^{\beta^*} = \frac{1}{1+{\beta}} \PhiT_f^{n^*} + \frac{{\beta}}{1+{\beta}} \PhiT_f^o.
\end{equation}
Specifically, the linearization of the convection term in (\ref{momentumeqdiscr01}) is achieved by computing the volumetric flux $\phi_f^{\beta^*}$ according to (\ref{beta2}). Note that in the segregated solution approach, the system of equations is solved multiple times within one time step. Therefore, we have $\PhiT_f^{n^*} = \PhiT_f^{o}$ when the equations are solved for the first time after the time index is increased by one. But as the number of outer iterations on the same time-level is increased, $\PhiT_f^{n^*}$ is updated with the latest values, subsequently to the solution of the respective equations, such that it converges successively towards $\PhiT_f^{n}$.

In the linearized convection term in (\ref{momentumeqdiscr01}), the face-centroid velocity $\velocity_f^{n}$ is handled implicitly using the high-resolution differencing scheme Gamma \cite{Jasak1999}. The Gamma scheme guarantees the boundedness of the solution by combining second-order accurate central differencing with unconditionally bounded upwind differencing. For the blending factor between the linear and the upwind scheme, we use the value $\beta_m = 1/10$, which is recommended in \cite{Jasak1999}, in order to preserve second-order accuracy and to avoid convergence problems.

The derivative $\vec{s}_f \dprod \left(\nabla {\velocity}\right)_f^{n}$ in the diffusion term of (\ref{momentumeqdiscr01}) is discretized according to \cite{Mathur1997} as
\begin{equation}
\label{laplacian}
\vec{s}_f \dprod \left(\nabla {\velocity}\right)_f^{n} 
= 
\underbrace{\frac{\left\vert \vec{s}_f \right\vert}{\vec{n}_f \dprod \vec{d}_f} \left(\velocity_N^n - \velocity_P^n \right)}_{\textnormal{orthogonal contribution}}
+ 
\underbrace{\left(\vec{s}_f - \frac{\left\vert \vec{s}_f \right\vert}{\vec{n}_f \dprod \vec{d}_f}\vec{d}_f \right) \dprod \left(\nabla \velocity \right)_f^{n^*}
}_{\textnormal{non-orthogonal correction}},
\end{equation}
where $\vec{d}_f = \vec{x}_N - \vec{x}_P$ represents the connection vector of the two computational points $P$ and $N$. The orthogonal contribution is handled implicitly by evaluating the current second-order accurate central difference of the neighboring cells directly at face $f$. The non-orthogonal correction term is evaluated explicitly by linear interpolation of the cell-centroid gradients at $P$ and $N$ to face $f$ according to
\begin{equation}
\label{linearinterpolation}
\left(\nabla \velocity \right)_f^{n^*} = \xi_f {\left(\nabla \velocity\right)}_P^{n^*} + \left(1 - \xi_f \right) {\left(\nabla \velocity\right)}_N^{n^*},
\end{equation}
where the linear interpolation factor to interpolate between the cell-centroids $P$ and $N$ is defined as $\xi_f = {\overline{f N}}/{\overline{P N}}$. In (\ref{linearinterpolation}), the cell-centroid gradients are computed using a discretized form of the Gauss' theorem so that for the centroid $P$ of a CV with $V_P$ we have
\begin{equation}
\label{gaussgrad01}
{\left(\nabla \velocity\right)}_P^{n^*} = \frac{1}{V_P}\sum_f \left(\vec{s}_f {\velocity}_f^{n^*}\right).
\end{equation}


In pressure-based solution approaches, the momentum equation is not necessarily solved. Instead, an additional pressure equation is constructed from the discretized continuity and momentum equations. The pressure-velocity coupling problem is then handled in a segregated iterative procedure by solving the pressure equation and subsequently updating the velocity, using the new pressure. In this work, the pressure-implicit with splitting of operators (PISO) algorithm \cite{Issa1986} is used to time-advance the pressure and the velocity. Detailed explanations of the PISO algorithm in OpenFOAM can be found, e.g.,\ in \cite{Darwish2015}. However, for the VoF solver interFoam, the pressure equation is constructed with a modified pressure defined by
\begin{align}
\label{modpressure}
p_d &= p - \rho \vec{b} \dprod \vec{x}_P,\\
\label{defpressured}
\nabla p_d &= \nabla p - \rho \vec{b} - \vec{b} \dprod \vec{x}_P \nabla \rho.
\end{align}
In (\ref{modpressure}), the hydrostatic pressure is removed from $p$ to yield the modified pressure $p_d$. Using the modified pressure as dependent variable is advantageous in a collocated variable arrangement with respect to the numerical treatment of density jumps at the interface. To avoid decoupling of the fields by an interpolation of the cell-centered density gradient term to the face, the density gradient is considered in the Rhie and Chow correction by a compact discretization directly at the cell-face. 
Substituting the pressure gradient in (\ref{momentumeqdiscr01}) with (\ref{defpressured}), the discretized momentum equation can be written as a linear algebraic equation as
\begin{align}
\label{algebraicform02}
a_P \velocity_P^n + \sum_N a_N \velocity_N^n 
=
{\vec{r}}_P 
- {\left(\nabla p_d\right)}_P^{n}
- \vec{b} \dprod \vec{x}_P {\left(\nabla \rho\right)}_P^{n^*} 
+ \sigma {\left({\smalloverbrace{\kappa}}\right)}_P^{n^*} {\left(\nabla \alpha_\phaseOne\right)}_P^{n^*}
+ {\left({\overline{\stressP}^\phaseOne}\right)}_P^{n^*} \dprod {\left(\nabla \alpha_\phaseOne\right)}_P^{n^*},
\end{align}
where the coefficients $a_P$ and $a_N$ read
\begin{align}
\label{coeffap01a}
a_P &= \frac{1 + \beta}{\vartriangle \! t} \rho + \frac{1}{{V_P}}\sum_f {\svisc}_f \frac{\left\vert \vec{s}_f \right\vert}{\vec{n}_f \dprod \vec{d}_f} + a_P^{c},\\
\label{coeffan02a}
a_N &= - \frac{{\svisc}_f}{{V_P}}  \frac{\left\vert \vec{s}_f \right\vert}{\vec{n}_f \dprod \vec{d}_f} + a_N^{c}.
\end{align}
The term ${\vec{r}}_P$ in (\ref{algebraicform02}) contains all explicitly evaluated lagged and old-time level terms, except for the other terms on the r.h.s.\ of (\ref{algebraicform02}), which are still formulated in a non-discretized form. 
The coefficients $a_P^{c}$ and $a_N^{c}$ in (\ref{coeffap01a}) and (\ref{coeffan02a}) represent the contributions from the discretization of the convection term, using the Gamma discretization scheme. For the purpose of the derivation of the pressure equation, $\nabla p_d^{n}$ must be considered at the new time-level $n$. Moreover, since the PISO algorithm is an iterative procedure in which the pressure and velocity are sequentially updated, we introduce another \mbox{symbol $'$} to denote lagged terms with respect to the outer PISO iterations. With the definition 
\begin{align}
\label{algebraicform03}
\vec{h}_{P} := {\vec{r}}_P - \sum_N a_N \velocity_N^{n'},
\end{align}
equation (\ref{algebraicform02}) can be rearranged to express the cell-centroid velocity as
\begin{align}
\label{algebraicform04}
\velocity_{P}^{n} = \frac{\vec{h}_{P}}{a_P} + \frac{1}{a_P} \left[- {\left(\nabla p_d\right)}_P^{n}
- \vec{b} \dprod \vec{x}_P {\left(\nabla \rho\right)}_P^{n^*} 
+ \sigma {\left({\smalloverbrace{\kappa}}\right)}_P^{n^*} {\left(\nabla \alpha_\phaseOne\right)}_P^{n^*}
+ {\left({\overline{\stressP}^\phaseOne}\right)}_P^{n^*} \dprod {\left(\nabla \alpha_\phaseOne\right)}_P^{n^*}\right].
\end{align}
This cell-centroid velocity is used to construct a cell-face velocity equation as
\begin{equation}
\label{vel03}
\velocity_f^{n} = \left(\frac{\vec{h}}{a}\right)_f + \left(\frac{1}{a}\right)_f 
\left[
-\left(\nabla p_d \right)_f^{n}
- \vec{b} \dprod \vec{x}_f {\left(\nabla \rho\right)}_f^{n^*} 
+ \sigma {\left({\smalloverbrace{\kappa}}\right)}_f^{n^*} {\left(\nabla \alpha_\phaseOne\right)}_f^{n^*}
+ {\left({\overline{\stressP}^\phaseOne}\right)}_f^{n^*} \dprod {\left(\nabla \alpha_\phaseOne\right)}_f^{n^*}\right]
,
\end{equation}
where $\left(\frac{\vec{h}}{a}\right)_f$ and $\left(\frac{1}{a}\right)_f$ are computed by linear interpolation (\ref{linearinterpolation}) of the cell-centroid expressions at $P$ and $N$, next to the face $f$. Inserting (\ref{vel03}) into the discretized continuity equation
\begin{equation}
\label{contidiscretization01}
\sum_{f} \vec{s}_f \dprod \velocity_f^{n} = 0,
\end{equation}
yields, after rearranging, the pressure equation
\begin{equation}
\label{contidiscretization02}
\sum_{f} \left(\frac{1}{a}\right)_f \vec{s}_f \dprod \left(\nabla p_d \right)_f^{n} = \sum_{f}
\left(\phi_f^{n'} + \phi_f^{n^*}\right)
,
\end{equation}
where the term $\vec{s}_f \dprod \left(\nabla p_d \right)_f^{n}$ is discretized analogous to (\ref{laplacian}) as
\begin{equation}
\label{laplacianpeq}
\vec{s}_f \dprod \left(\nabla p_d\right)_f^{n} 
= 
\underbrace{\frac{\left\vert \vec{s}_f \right\vert}{\vec{n}_f \dprod \vec{d}_f} \left(\left({p_d}\right)_N^n - \left({p_d}\right)_P^n \right)}_{\textnormal{orthogonal contribution}}
+ 
\underbrace{\left(\vec{s}_f - \frac{\left\vert \vec{s}_f \right\vert}{\vec{n}_f \dprod \vec{d}_f}\vec{d}_f \right) \dprod \left(\nabla {p_d} \right)_f^{n'}
}_{\textnormal{non-orthogonal correction}}.
\end{equation}
The cell-face fluxes $\phi_f^{n'}$ and $\phi_f^{n^*}$ in (\ref{contidiscretization02}) read
\begin{align}
\label{fluxp01}
\phi_f^{n'}
&=
\vec{s}_f \dprod \left(\frac{\vec{h}}{a}\right)_f, \\
\label{fluxp02}
\phi_f^{n^*}
&= 
\left(\frac{1}{a}\right)_f 
\left[
%
- \gamma_{b} \vec{s}_f \dprod  {\left(\nabla \rho\right)}_f^{n^*}
%
+ \gamma_{\sigma \kappa} \vec{s}_f \dprod {\left(\nabla \alpha_\phaseOne\right)}_f^{n^*}
+ \widehat{\vec{s}}_f^{n^*} \dprod {\left(\nabla \alpha_\phaseOne\right)}_f^{n^*}
\right].
\end{align}
In (\ref{fluxp02}), the scalar coefficient in the body force reads $\gamma_{b} = \vec{b} \dprod \vec{x}_f$, the coefficient in the surface tension force reads
\begin{equation}
\gamma_{\sigma \kappa} = \sigma {\left({\smalloverbrace{\kappa}}\right)}_f^{n^*},
\end{equation}
where the curvature is obtained according to the CSF model (\ref{stf01}) by linear interpolation of the cell-centered values, i.e.\
\begin{equation}
{\left({\smalloverbrace{\kappa}}\right)}_P^{n^*} = -\frac{1}{V_P} \sum_f \vec{s}_f \dprod \left(\frac{{\left(\nabla \alpha_\phaseOne\right)}_f^{n^*}}{\vert {\left(\nabla \alpha_\phaseOne\right)}_f^{n^*} \vert}\right),
\end{equation}
to the cell-face. The modified vector $\widehat{\vec{s}}_f^{n^*}$ in the interface stress term in (\ref{fluxp02}) reads
\begin{equation}
\label{mods}
\widehat{\vec{s}}_f^{n^*} =
\vec{s}_f \dprod {\left({\overline{\stressP}^\phaseOne}\right)}_f^{n^*}.
\end{equation}
With (\ref{mods}),
we find 
\begin{equation}
\label{istressdiscrete0}
\widehat{\vec{s}}_f^{n^*} \dprod {\left(\nabla \alpha_\phaseOne\right)}_f^{n^*}
= 
{\left\vert \widehat{\vec{s}}_f^{n^*} \right\vert} {\vec{n}}_f \dprod {\left(\nabla \alpha_\phaseOne\right)}_f^{n^*}
+
\left(\widehat{\vec{s}}_f^{n^*}  - {\left\vert \widehat{\vec{s}}_f^{n^*} \right\vert} {\vec{n}}_f\right) \dprod {\left(\nabla \alpha_\phaseOne\right)}_f^{n^*}.
\end{equation}
Inserting a compact discretization of the volume-fraction gradient at the cell-face as
\begin{equation}
\label{istdiscrete1}
\vec{n}_f \dprod {\left(\nabla \alpha_\phaseOne\right)}_f^{n^*}
= 
{\frac{1}{\vec{n}_f \dprod \vec{d}_f} \left(\left({\alpha_\phaseOne}\right)_N^{n^*} - \left({\alpha_\phaseOne}\right)_P^{n^*} \right)}
+
{\left({\vec{n}}_f - \frac{1}{\vec{n}_f \dprod \vec{d}_f}\vec{d}_f \right) \dprod \left(\nabla {\alpha_\phaseOne} \right)_f^{n^*}
}
\end{equation}
into (\ref{istressdiscrete0}), we obtain the new scheme for the interface stress term as
\begin{equation}
\label{istressdiscrete}
\widehat{\vec{s}}_f^{n^*} \dprod {\left(\nabla \alpha_\phaseOne\right)}_f^{n^*}
= 
{\frac{\left\vert \widehat{\vec{s}}_f^{n^*} \right\vert}{{\vec{n}}_f \dprod \vec{d}_f} \left(\left({\alpha_\phaseOne}\right)_N^{n^*} - \left({\alpha_\phaseOne}\right)_P^{n^*} \right)}
+ 
{\left(\widehat{\vec{s}}_f^{n^*} - \frac{\left\vert \widehat{\vec{s}}_f^{n^*} \right\vert}{{\vec{n}}_f \dprod \vec{d}_f}\vec{d}_f \right) \dprod \left(\nabla {\alpha_\phaseOne} \right)_f^{n^*}
}.
\end{equation}
Similar to (\ref{laplacian}), the first term on the r.h.s.\ of (\ref{istressdiscrete}) is computed directly at the cell-face $f$ as the second-order accurate central difference between the two neighboring cell-centroids, while the second term on the r.h.s.\ of (\ref{istressdiscrete}) represents the correction, which is obtained by interpolating the cell-centroid gradients to the cell-face. 
With (\ref{istressdiscrete}) it becomes obvious that the discretization of the interface stress term is computed on the same computational stencil as the surface tension force term. Both contributions can be combined to yield
\begin{equation}
\label{istdiscretec01}
\left(\gamma_{\sigma \kappa} \vec{s}_f + \widehat{\vec{s}}_f^{n^*}\right) \dprod {\left(\nabla \alpha_\phaseOne\right)}_f^{n^*}
= 
\gamma_{\perp}
{ \left(\left({\alpha_\phaseOne}\right)_N^{n^*} - \left({\alpha_\phaseOne}\right)_P^{n^*} \right)}
+
{
\vec{k}
\dprod \left(\nabla {\alpha_\phaseOne} \right)_f^{n^*}
},
\end{equation}
where
\begin{align}
\label{istdiscretec02}
\gamma_{\perp} &= 
{\frac{
\gamma_{\sigma \kappa} \left\vert {\vec{s}}_f \right\vert
+
\left\vert \widehat{\vec{s}}_f^{n^*} \right\vert
}
{\vec{n}_f \dprod \vec{d}_f}}
\end{align}
and
\begin{align}
\label{istdiscretec03}
\vec{k} &=
{
{\gamma_{\sigma \kappa} {\vec{s}}_f}
+ {\widehat{\vec{s}}_f^{n^*}}
- {\frac{\gamma_{\sigma \kappa}{\left\vert {\vec{s}}_f \right\vert} + {\left\vert \widehat{\vec{s}}_f^{n^*} \right\vert}}{\vec{n}_f \dprod \vec{d}_f}\vec{d}_f }
}.
\end{align}
This completes the discretization of the fluxes in the pressure equation (\ref{contidiscretization02}) on general unstructured meshes. 
After solving the pressure equation (\ref{contidiscretization02}), the reconstructed cell-face flux
\begin{equation}
\label{contidiscretization03}
\phi_f^n = \phi_f^{n'} + \phi_f^{n^*} - \left(\frac{1}{a}\right)_f \vec{s}_f \dprod \left(\nabla p_d \right)_f^{n}
\end{equation}
is conservative, which means that
\begin{equation}
\label{contidiscretization04}
\sum_{f} \phi_f^n = 0.
\end{equation}
The discretization of the constitutive equation can be carried out in the same way as proposed in \cite{Niethammer2017} for a single-phase system. Defining the generic working variable $\GT := \cvec{F}(\overline{\CT}^\phaseOne)$ and applying the discretization described above to the volume-averaged constitutive equation (\ref{constCEqva03}), we have
\begin{equation}
\label{constdiscretization01}
\frac{V_P}{\vartriangle \! t} \left( \GT_P^n - \GT_P^o \right) + \sum_{f} \phi_f^{\beta^*} \GT_f^{\beta} = V_P \cvec{Q}_{\GT_P}^{\beta^*},
\end{equation}
where $\cvec{Q}_{\GT_P}^{\beta^*}$ represents the explicitly evaluated source terms.
The corresponding linear algebraic equation has the form
\begin{equation}
\label{algebraicformG02}
a_P \GT_P^n + \sum_N a_N \GT_N^n = \RT_P,
\end{equation}
where the coefficients $a_P$ and $a_N$ read
\begin{align}
\label{coeffap01b}
a_P &= \frac{1 + \beta}{\vartriangle \! t} V_P + a_P^{c},\\
\label{coeffan02b}
a_N &= a_N^{c}
\end{align}
and where the symmetric tensor $\RT_P$ contains the remaining old-time level terms and the lagged terms.

\subsection{Interface capturing}
\label{subsec:intcapt}
An algebraic VoF method is used to capture the interface. In such a method, the interface representation is known implicitly from the distribution of the volume fraction in the computational domain. A geometrical reconstruction of the interface is not required, which dramatically reduces the difficulty of implementation in multidimensions on general unstructured computational grids \cite{Jofre2014, Maric2018}. The interface is advected by solving the mixture model transport equation for the volume fraction (\ref{vofraceq02b}). Thus, the accuracy of the interface representation and transport depends strongly on the numerical differencing scheme applied to the advection term in (\ref{vofraceq02b}). The high accuracy numerical differencing scheme is required to simultaneously maintain sharpness of the interface, boundedness and conservation of the phase fraction. For the advection of a sharp profile it is well-known that higher-order schemes suffer from a lack of boundedness, i.e.\ they give rise to over- and undershoots, leading to unphysical results \cite{Waterson2007}. 
Several procedures have been developed to avoid such errors, which can be divided in two categories. The VoF-based interface capturing schemes apply a flux-limiter to the cell-face fluxes that enforces a one-dimensional monotonicity criterium. Widely used high resolution interface capturing schemes are the CICSAM scheme \cite{Ubbink1997, Ubbink1999} and the HRIC scheme \cite{Muzaferija1998}. Both schemes are based on the normalized variable formulation \cite{Leonard1988a}, together with the convection boundedness criterion \cite{Gaskell1988}. Such schemes can be applied in complex two- and three-dimensional flows on general unstructured computational grids, however they neither guarantee the solution to remain sharp nor bounded in this case. An alternative approach which is used here is the multi-dimensional FCT method, introduced by Boris and Book \cite{Boris1973} and extended to multiple dimensions by Zalesak \cite{Zalesak1979}. The FCT method is a fully multi-dimensional flux limiting strategy that applies a local weighting-procedure between an unconditionally bounded low order upwind differencing scheme and an anti-diffusive higher order correction scheme, such that the higher order flux is used to the greatest extent possible to keep the solution bounded between some local extrema.
Given the time-explicit discretization of the volume fraction transport equation (\ref{vofraceq02b}), i.e.\
\begin{align}
\label{vofraceq03}
\frac{ V_P}{\vartriangle \! t} \left(\alpha_\phaseOne^n - \alpha_\phaseOne^o\right)
+ \sum_f \phi_f^{\beta^*}
= 0,
\end{align}
where the flux $\phi_f$ at the cell-face $f$ is defined as $\phi_f = \vec{s}_f \dprod {\velocity}_f {\alpha_\phaseOne}_f$, the FCT method introduces a blending between a lower-order flux $\phi_{f,\;\! L}$ and a higher-order flux $\phi_{f,\;\! H}$ as
\begin{align}
\label{vofraceq04}
\frac{ V_P}{\vartriangle \! t} \left(\alpha_\phaseOne^n - \alpha_\phaseOne^o\right)
+ \sum_f \phi_{f,\;\! L}^{\beta^*} + \sum_f \lambda_f \left(\phi_{f,\;\! H}^{\beta^*} - \phi_{f,\;\! L}^{\beta^*}\right)
= 0,
\end{align}
where $\lambda_f$ is the flux limiter defined on each cell-face $f$. Here, $\lambda_f$ has to be determined in an appropriate manner to ensure that the higher-order fluxes do not introduce any new local extrema into the solution. The bounds are estimated for each CV from a local stencil at the point $P$ and the CVs $n$ cell-face neighbors as
\begin{align}
\label{vofraceq05a}
\alpha_{\phaseOne,\:\!P,\:\! \operatorname{min}} &:= {\operatorname{min}} \left(\alpha_{\phaseOne,\:\! P}^{\beta^*},\:\! \alpha_{\phaseOne,\:\! N,\:\! 1}^{\beta^*} ,\:\!...,\:\! \alpha_{\phaseOne,\:\! N,\:\! n}^{\beta^*}\right),\\
\label{vofraceq05b}
\alpha_{\phaseOne,\:\!P,\:\! \operatorname{max}} &:= {\operatorname{max}} \left(\alpha_{\phaseOne,\:\! P}^{\beta^*},\:\! \alpha_{\phaseOne,\:\! N,\:\! 1}^{\beta^*},\:\!...,\:\!  \alpha_{\phaseOne,\:\! N,\:\! n}^{\beta^*}\right).
\end{align}
Then, to satisfy the condition $\alpha_{\phaseOne,\:\! \operatorname{min}} \leq \alpha_{\phaseOne}^n \leq \alpha_{\phaseOne,\:\! \operatorname{max}}$, a cell-based limiter can be obtained. Defining the correction $\phi_{f,\;\! \textnormal{corr}}^{\beta^*} := \phi_{f,\;\! H}^{\beta^*} - \phi_{f,\;\! L}^{\beta^*}$ in (\ref{vofraceq04}) and considering the fluxes \textit{into} the CV, we define
\begin{align}
\label{vofraceq06a}
P^+ &= -\sum_f \operatorname{min} \left( 0, \phi_{f,\;\! \textnormal{corr}}^{\beta^*} \right),\\
\label{vofraceq06b}
Q^+ &= \frac{V_P}{\vartriangle \! t} \left(\alpha_{\phaseOne,\:\!P,\:\! \operatorname{max}} - \alpha_\phaseOne^o\right) + \sum_f \phi_{f,\;\! L}^{\beta^*},\\
\label{vofraceq06c}
R^+ &= -\sum_f \operatorname{min} \left( 0, \lambda_f \phi_{f,\;\! \textnormal{corr}}^{\beta^*} \right).
\end{align}
Similarly for the fluxes \textit{away} from the CV, we have
\begin{align}
\label{vofraceq06d}
P^- &= \sum_f \operatorname{max} \left( 0, \phi_{f,\;\! \textnormal{corr}}^{\beta^*} \right),\\
\label{vofraceq06e}
Q^- &= \frac{V_P}{\vartriangle \! t} \left(\alpha_{\alpha_\phaseOne^o - \phaseOne,\:\!P,\:\! \operatorname{min}} \right) - \sum_f \phi_{f,\;\! L}^{\beta^*},\\
\label{vofraceq06f}
R^- &= \sum_f \operatorname{max} \left( 0, \lambda_f \phi_{f,\;\! \textnormal{corr}}^{\beta^*} \right).
\end{align}
With the definitions (\ref{vofraceq06a}) to (\ref{vofraceq06f}), the least upper and greatest lower bound on the fraction, which must multiply all higher order fluxes, is then given by
\begin{align}
\label{vofraceq06g}
S^{\pm} &= \operatorname{max} \left[ 0, \operatorname{min} \left( 1, \frac{Q^\pm + R^\mp}{P^\pm} \right) \right].
\end{align}
Note that the choice (\ref{vofraceq06g}) is slightly different from Zalesak's \cite{Zalesak1979}, since it additionally takes into account the higher-order net flux of the CV. The cell-face limiter $\lambda_f$ is obtained from the bounds at the points $P$ and $N$ on either side of the face $f$ as
\def\arraystretch{1.4}
\begin{align}
\label{vofraceq06h}
\lambda_f = 
\left\{
\begin{array}{ll}
\operatorname{min} \left(S_P^-,\;\! S_N^+\right) \quad \textnormal{if} \ \phi_{f,\;\! \textnormal{corr}}^{\beta^*} > 0,\\
\operatorname{min} \left(S_P^+,\;\! S_N^- \right) \quad \textnormal{if} \ \phi_{f,\;\! \textnormal{corr}}^{\beta^*} < 0.
\end{array}
\right.
\end{align}
The FCT limiter estimation is carried out with the so-called MULES algorithm in OpenFOAM. Unlike Zalesak's FCT \cite{Zalesak1979}, the MULES algorithm applies an iterative strategy for the estimation of $\lambda_f$. Initially, the limiter is computed with (\ref{vofraceq05a}) to (\ref{vofraceq06h}), using an initial guess of $\lambda_f = 1$. Then (\ref{vofraceq06c}) and (\ref{vofraceq06f}) to (\ref{vofraceq06h}) are computed again with the result for $\lambda_f$ from the previous iteration. In total two additional iterations are applied in this work.

While the FCT method is powerful to preserve the boundedness of the solution in multi-dimensional flows on general unstructured meshes, the sharpness of the interface is not inherently guaranteed. The latter aspect primarily depends on how the cell-face fluxes $\phi_{f,\;\! L}^{\beta^*}$ and $\phi_{f,\;\! H}^{\beta^*}$ are computed. The most natural choice is to apply a combination of upwind differencing for $\phi_{f,\;\! L}^{\beta^*}$ and downwind differencing for $\phi_{f,\;\! H}^{\beta^*}$, which will result in a compressive overall scheme in (\ref{vofraceq04}). Downwind differencing will introduce anti-diffusion to counteract the numerical diffusion of the first-order upwind scheme, applied for $\phi_{f,\;\! L}^{\beta^*}$.  
For the higher order flux $\phi_{f,\;\! H}^{\beta^*}$, we apply the inter-Gamma scheme \cite{Jasak1995}, which is a reasonable choice since it uses a certain amount of downwind differencing and shows good shape preserving properties at small CFL numbers \cite{Gopala2008}. The inter-Gamma scheme is based on the NVD approach. As such, it is subject to similar issues as mentioned above if it is used for multi-dimensional flows on general unstructured computational grids. Under certain conditions, it might fail to keep the interface sharp. One possible source of errors is related to the fact that the scheme's compression is completely governed by the one-directional convective fluxes through the cell-faces of the CV. As a consequence, the scheme's interface compression is not necessarily directed bi-normal to the interface orientation, unless the scheme's geometric weights are modified to account for the local interface orientation at the cell-face. However, even with a modification of the geometric weights the interface compression of the inter-Gamma scheme can locally become too small in magnitude to balance the numerical diffusion of the interface. 
To circumvent such errors, we use an alternative approach in OpenFOAM which combines the higher-order flux in (\ref{vofraceq03}) with an extra compressive flux as
\begin{align}
\label{icm01}
\frac{ V_P}{\vartriangle \! t} \left(\alpha_\phaseOne^n - \alpha_\phaseOne^o\right)
+ \sum_f \phi_f^{\beta^*} + \sum_f \phi_{f,\;\! c}^{\beta^*} \alpha_{f, \phaseOne}^{\beta^*}(1 - \alpha_{f, \phaseOne}^{\beta^*})
= 0.
\end{align}
This is sometimes referred to as the interface compression scheme by Weller \cite{Rusche2002}. As can be seen from (\ref{icm01}) it is formally equivalent to the addition of a ``counter-gradient'' transport term to the volume fraction equation. Containing the factor $\alpha_{f, \phaseOne}^{\beta^*}(1 - \alpha_{f, \phaseOne}^{\beta^*})$, this extra term is only active in the interface region. One advantage of the method is that the compressive flux $\phi_{f,\;\! c}^{\beta^*}$ is computed such that it is acting in the direction of the interface unit normal vector $\vec{n}_{\Sigma, f} = {{\left(\nabla \alpha_\phaseOne\right)}_f}/{\vert {\left(\nabla \alpha_\phaseOne\right)}_f \vert}$. However, there is no physical justification for the appearance of this extra flux, it is motivated by purely numerical considerations, i.e.\ to sharpen the interface by bi-normal interface compression. In this work $\phi_{f,\;\! c}^{\beta^*}$ is modeled by
\begin{align}
\label{icm02}
\phi_{f,\;\! c}^{\beta^*} = \operatorname{min} \left(c_\alpha \frac{\left\vert\phi_{f}^{\beta^*}\right\vert}{\vert \vec{s}_f \vert}, \operatorname{max}\left[\frac{\left\vert\phi_{f}^{\beta^*}\right\vert}{\vert \vec{s}_f \vert}\right]\right)\left(\vec{n}_{\Sigma, f} \dprod \vec{s}_f \right).
\end{align}
The idea behind (\ref{icm02}) is that the magnitude of the cell-face flux $\phi_{f}^{\beta^*}$ is used as a local measure for the interface diffusion speed which is being counteracted by $\phi_{f,\;\! c}^{\beta^*}$. The model coefficient $c_\alpha$ controls the amount of interface compression and is chosen to be $1$.

\subsection{Moving reference frame}
To save computational costs we use a local adaptive mesh refinement with a finer spatial resolution in the neighborhood of the fluid interface and a coarser resolution far away from the bubble. In an inertial frame of reference, the local adaptive mesh refinement needs to be adjusted during the simulation as the bubble moves inside the domain. Such a re-meshing procedure causes additional workload. When transforming the system into a moving reference frame that is attached to the motion of the rising bubble, re-meshing can be avoided. In the moving reference frame, additional forces will appear due to the different acceleration of the bubble observed in both frames. Denoting the frame acceleration with $\vec{a}_F$, the body force in the non-inertial frame becomes
\begin{align}
{\rho}_m \vec{b} = {\rho}_m \left(\vec{g} - \vec{a}_F\right),
\end{align}
where the $\vec{g}$ is the gravitational acceleration and the frame acceleration $\vec{a}_F$ is given by
\begin{align}
\vec{a}_F = \frac{d {\velocity}_F}{d t}.
\end{align}
The time-dependent velocity ${\velocity}_F$ is also prescribed at the inlet of the computational domain. The velocity ${\velocity}_F$ is computed in every time step of the simulation, based on the relative movement of the center of mass of the bubble. To dampen small oscillations in the velocity field a PD controller is applied for the computation of ${\velocity}_F$. 


\section{Results}
Numerical results for a bubble rising in a viscoelastic fluid are presented in this section. A particular focus is put on the study of characteristic effects of the rising bubble, such as the jump discontinuity of the steady state rise velocity and the negative wake behind the trailing end of the bubble. The major goal is to achieve quantitative agreement of the bubble rise velocities between the numerical results and experimental measurements for both subcritical and supercritical volumes. Furthermore, the local flow patterns around the bubble are analyzed to obtain a better knowledge of the non-trivial relations between the bubble rise velocity, the bubble shape, and the viscoelastic stress.

\subsection{Experiments, case setup and boundary conditions}
In the experiments by Pilz and Brenn, the rise of individual air bubbles in viscoelastic polymer solutions was investigated \cite{Pilz2007}. The aqueous polymer solutions were prepared using de-ionised water, stirring for several hours, allowing the solutions to rest for 24 hours, and taking every care to avoid that surfactants entered the liquids. The liquid was contained in a $450 \ \textnormal{mm}$ high cylindrical vessel with square-shaped cross section of $120 \ \textnormal{mm}$ side length. The bubbles were formed by accumulating below a spoon-like concave body in the liquid varying numbers of approximately $2 \ \textnormal{mm}^3$ bubbles produced from the end of a $0.09 \ \textnormal{mm}$ capillary by sucking air into the liquid through the capillary. The volume of the bubbles to be investigated was varied by this technique. The bubbles released from the spoon were allowed to rise for some 15 to 35 bubble diameters before entering the containment. The bubble rise velocity measurement was carried out $350 \ \textnormal{mm}$ above the bottom of the vessel, where the bubbles were at their state of steady rise and shape. The bubble volume and velocity were measured based on processing of bubble images acquired with a video camera at the framing rate of $30 \ \textnormal{Hz}$. Bubble volumes ranged between $3 \ \textnormal{mm}^3$ and $990 \ \textnormal{mm}^3$. Velocities were of the order of some $\textnormal{mm}/\textnormal{s}$ to some $100 \ \textnormal{mm}/\textnormal{s}$, so that continuous illumination with a mercury lamp was appropriate for producing sharp bubble images. The aqueous polymer solutions were prepared using de-ionised water, taking every care to avoid that surfactants entered the liquids.

The three-dimensional computational domain used in this study consists of a cube. The dimensions of the computational domain are adapted to the respective size of the bubble. In each case, the domain length is $21.7$ times larger than the initial bubble diameter. 
The simulation is initiated by placing a spherical bubble in the center of the domain.
Because the system is solved in a moving reference frame, the velocity boundary conditions of the computational domain need to be adjusted relative to the motion of the reference frame in each timestep. This is achieved by defining a Dirichlet boundary condition for the velocity on each side of the domain. This Dirichlet boundary condition is set with the time-dependent frame velocity ${\velocity}_F$. In every time step of the simulation, the boundary values are updated. For the pressure, the polymer stress and the volumetric phase fraction, we apply zero-gradient boundary conditions. Only on the horizontal bottom side of the cube different boundary conditions are used. Here, we expect an outflow. On the bottom side we apply zero-gradient boundary conditions for the velocity, the polymer stress and the volumetric phase fraction. For the pressure a Dirichlet boundary condition is used with a constant value.

\subsection{Fluid rheology}
\begin{figure}[htbp!]
\centering{
\includegraphics[width=420pt]{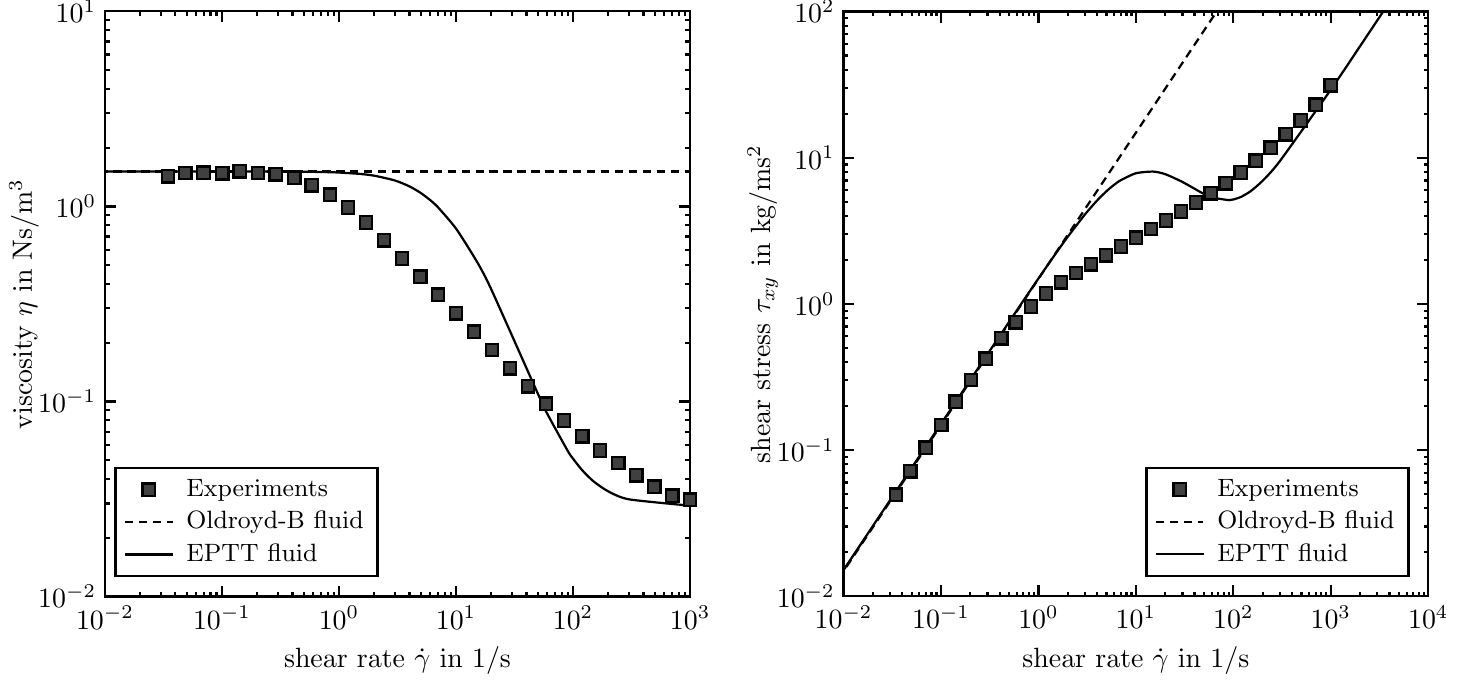}
}
\caption{\small Flow curves for the aqueous $0.8$ $\%$ weight P2500 solution. The symbols show the measurements in simple shear flow \cite{Pilz2007}. The curves show the viscosity (l.h.s.) and the shear stress (r.h.s.), using the Oldroyd-B and the EPTT model with the parameters given in Table \ref{tab:parameters}. Note that in the constitutive equations the relaxation time is not adjusted to the shear data; it is obtained from the measurements in elongational flow instead. For this reason, little freedom is left to shift the curve of the EPTT equation along the shear-rate axis.}
\label{fig:viscosityfit}
\end{figure}

The aqueous polymer solution used in the present work was characterized by Pilz and Brenn \cite{Pilz2007} by means of shear and elongational rheometry. Elongational rheometry was used for the determination of a characteristic relaxation time for the system, while the shear data was used to fit the viscosity. This procedure has been shown to result in a reliable estimation of the parameters for the constitutive equations presented in Table \ref{tab:constitutiveEx}, to model the physics of the rising bubble in the polymer solution. For the numerical experiments of the present work, we use the values of the relaxation time and the zero-shear viscosity proposed in \cite{Pilz2007}. Let us note that because of the solvent-polymer stress splitting, the total viscosity is decomposed as $\eta_0 = \svisc + \pvisc$. Depending on the rheological model, further parameters need to be estimated. The flow curve of the polymer solution in simple shear flow is shown in Figure \ref{fig:viscosityfit}. From this curve it can be seen that the polymer solution is shear-thinning. This behavior cannot be captured with simple two-parameter constitutive equations, like the Oldroyd-B model. Hence, we use the four-parameter exponential Phan-Thien Tanner (EPTT) model, which is frequently used in the literature for viscoelastic polymer solutions with shear-rate dependent flow behavior. A parameter fit of both models in simple shear flow is shown in Figure \ref{fig:viscosityfit}. An overview of all parameters is given in Table \ref{tab:parameters}.\!\!\!\!
\begin{table}[h!]
\small
\begin{center}
\begin{tabular}{@{}llllllll@{}}
\toprule
$w [\%]$ \ & $\rho_\phaseOne [\textnormal{kg}/\textnormal{m}^3]$ \ & $\pvisc [\textnormal{Ns}/\textnormal{m}^2]$ \ & $\svisc [\textnormal{Ns}/\textnormal{m}^2]$ \ & $\tau [\textnormal{s}]$ \ & $\sigma [\textnormal{N}/\textnormal{m}]$ \ & $\epsilon$ \ & $\zeta$ \ \\ 
\midrule
$0.8$ & $1000.9$ & $1.483$ & $0.03$ & $0.203$ & $0.07555$ & $0.05$ & $0.12$ $\!$ \\
\bottomrule
\end{tabular}
\end{center}
\caption{\small Parameters for the polymer solution used for the investigation.}
\label{tab:parameters}
\end{table}\
\
The Oldroyd-B model is obviously not shear-rate dependent but exhibits a constant total viscosity over $\dot{\gamma}$. The EPTT model shows a shear-thinning behavior, which better corresponds to the measurements. Let us note that a closer fit to the experimental flow curve could be obtained by changing the relaxation time, which is what we want to avoid, because we trust the elongational data used to estimate $\tau$. Thus, we adjust only the parameters $\epsilon$, $\zeta$ and the viscosity ratio $\svisc/\eta_0$, while keeping $\tau$ unchanged. In simple shear flow, a change of $\epsilon$ and $\zeta$ results in a different slope of the viscosity curve, but does not shift the curve along the shear-rate coordinate. Hence, when keeping $\tau$ unchanged, the viscosity curve must have a steeper slope to achieve good agreement with the experiments.

Regarding the flow curves in Figure \ref{fig:viscosityfit}, an important question is how shear-thinning affects the velocity jump discontinuity. Results from other researchers \cite{Pillapakkam2007} suggest that the velocity jump occurs also for rheological models with shear-rate independent viscosities. This means that shear-thinning cannot be a necessary condition for the appearance of the velocity jump discontinuity. In the preparation of this work we made similar observations, using the Oldroyd-B equations. However, because the experiments unquestionably show that the polymer solution is shear-thinning, we use the EPTT equations for all numerical simulations of this study.

\subsection{Convergence of the bubble rise velocity}
We study the dependence of the numerical results for the bubble rise velocity on the spatial resolution of the computational mesh and the time step. To demonstrate that the grid resolution is sufficient to obtain mesh-independent rise velocities, we compare the numerical results of two different meshes. The mesh M1 is aimed to be used for the productive simulations of this work. We compare the rise velocities obtained on the mesh M1 with the results obtained on a coarser mesh M2. If the values for the rise velocities on the two meshes do not differ considerably, we consider the desired mesh M1 as sufficiently well resolved to achieve convergence in the rise velocities. A similar procedure is carried out regarding the time step size.

\begin{figure}[h]
\centering{
\includegraphics[width=399pt]{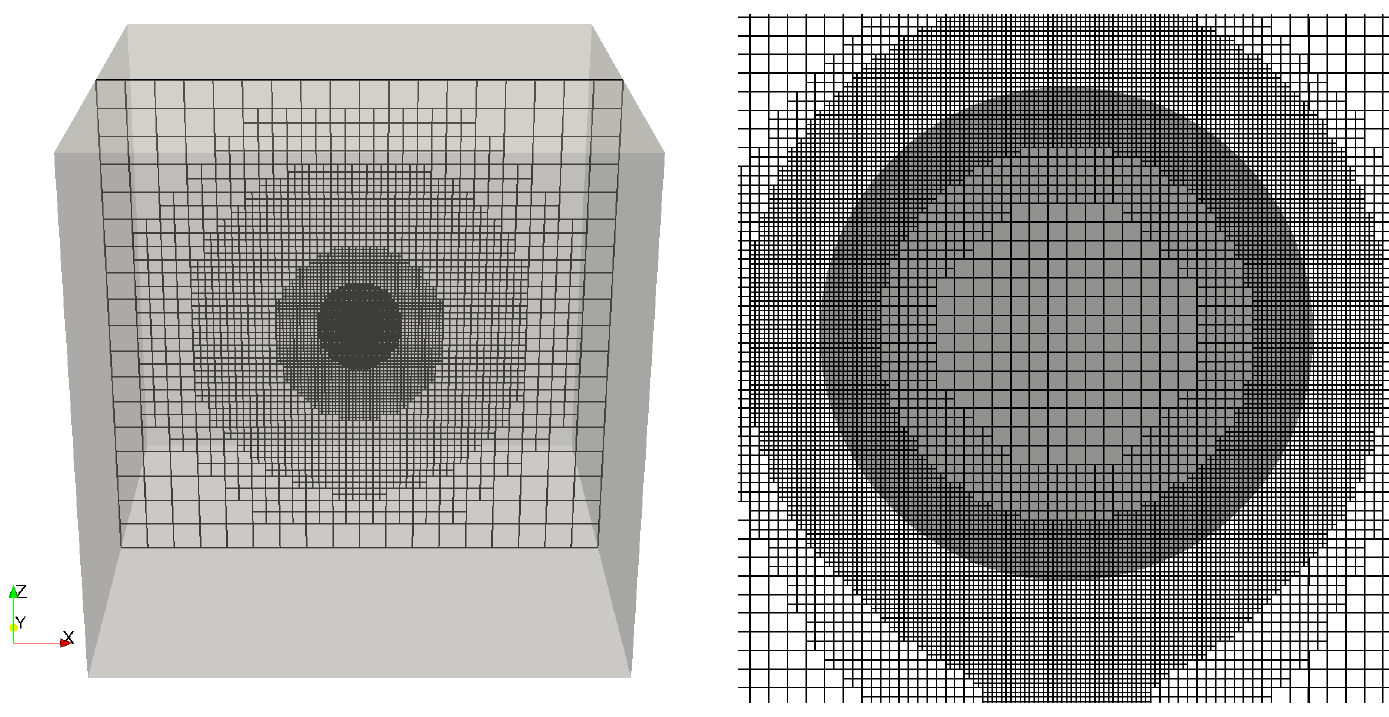}
}
\caption{\small Visualization of the three-dimensional computational mesh M1. Left: Cutting planar mesh slice through the center of the computational domain. Right: Detail of the mesh in the vicinity of the bubble. The initial bubble shape is represented in gray.}
\label{fig:meshdetail}
\end{figure}

The parameters of the two different meshes are shown in Table \ref{tab:meshes}.
\begin{table}[h!]
\small
\begin{center}
\begin{tabular}{@{}lllll@{}}
\toprule
Name \ & $n_{\textnormal{CV}}$ \ & \ $n_{\textnormal{CV},\operatorname{min}}/d_b$ & $l_{\textnormal{CV},\operatorname{min}} / d_b$ \ & $L/d_b\:$\\
\midrule
M1 \ & $1577864$ \ & $108$ \ & $9.42 \times 10^{-3}$ \ & $21.7$ \\
M2 \ & $790300$ \ & $84$ \ & $1.19 \times 10^{-2}$ \ & $21.7$ \\
\bottomrule
\end{tabular}
\end{center}
\caption{\small Mesh characteristics. Number of control volumes $n_{\textnormal{CV}}$ of the mesh, ratio of the number of control volumes per bubble diameter $n_{\textnormal{CV},\operatorname{min}}/d_b$ on the finest refinement level, minimum length of the CV per bubble diameter $l_{\textnormal{CV},\operatorname{min}} / d_b$ on the finest refinement level and ratio of the domain length per bubble diameter $L/d_b$.}
\label{tab:meshes}
\end{table}
Local adaptive mesh refinement is used to obtain high spatial resolution in the vicinity of the bubble interface and to save computational resources in less important regions far away from the bubble, by an increased mesh-spacing. Details of the computational meshes are shown in Figure \ref{fig:meshdetail}.

The fine mesh M1 has $\approx 1.58$ million computational CVs and a ratio of $108$ CVs per initial bubble diameter on the finest refinement level. In the coarse mesh M2 the total number of cells is approximately reduced by half, while keeping the local distribution of the mesh refinement zones unchanged. The ratio of CVs per bubble diameter is reduced to $84$ in M2. For both meshes the domain length is $21.7$ times larger than the bubble diameter. This relatively large $L/d_b$ ratio is sufficient to eliminate wall effects in the flow field around the bubble. In order to show that the numerical results are independent from the time step size, we compare the numerical results, obtained with a constant time step of $2\times 10^{-6}$ seconds with the results, obtained with twice this time step size. Note that the time step sizes in our simulations are specified, depending on the fluid relaxation time. An additional criterion makes sure that the CFL number stays below the value $0.25$. The time step $2\times 10^{-6}$ seconds, used in the present investigations, is small enough to fulfill the CFL criterion in every time step throughout the simulation, such that the time step size never falls below $2\times 10^{-6}$ seconds.

\begin{figure}[h]
\centering{
\includegraphics[width=398pt]{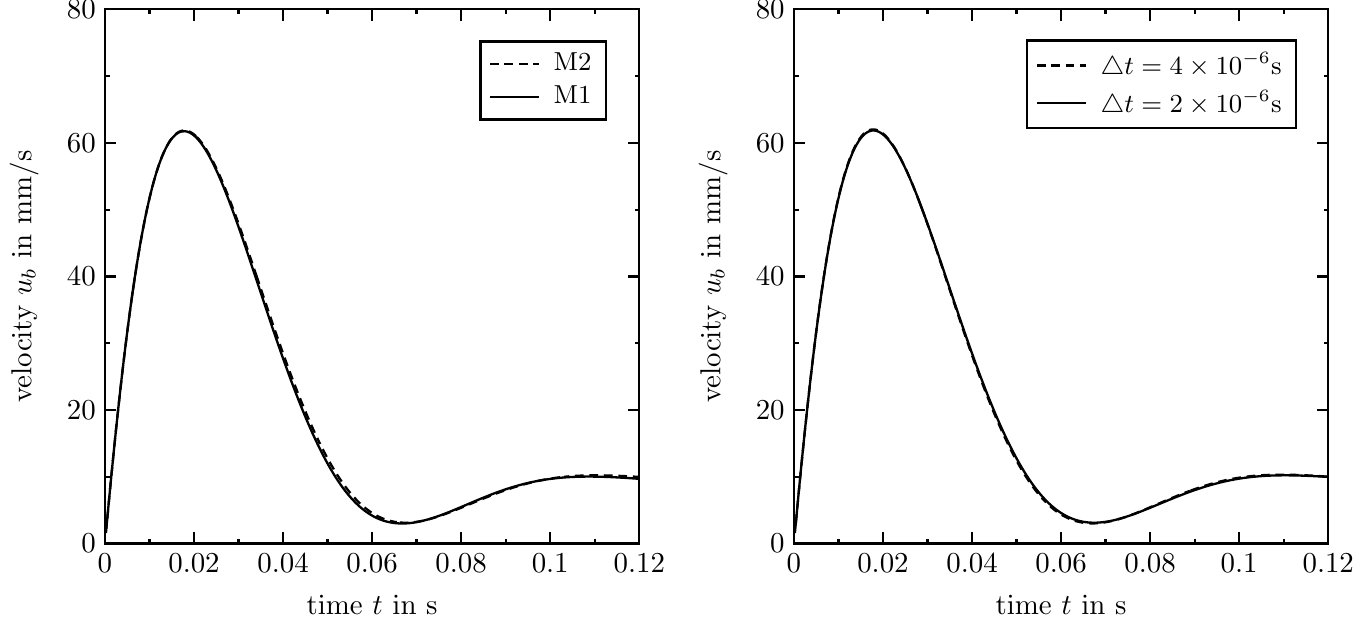}
}
\caption{\small Transient bubble rise velocity as a function of time of a bubble accelerating from rest in a viscoelastic fluid. Left: Convergence with mesh resolution, using the time-step size $\;\triangle t = 2 \times 10^{-6} s$. Right: Convergence with the time step, using the mesh M2.}
\label{fig:convergencesub}
\end{figure}

Figure \ref{fig:convergencesub} shows the transient bubble rise velocity as a function of time for a bubble accelerating from rest in a quiescent viscoelastic fluid.

On the l.h.s.\ two numerical simulations with similar settings were carried out on mesh M1 and mesh M2, respectively. The deviation between the results obtained with the two different mesh resolutions is relatively small, such that we consider the fine mesh M1 as sufficiently well resolved to obtain convergence in the bubble rise velocities. All subsequent results are therefore obtained on the fine mesh M1. On the r.h.s.\ of Figure \ref{fig:convergencesub}, the transient rise velocity over time is shown for two different time step sizes, $2\times 10^{-6}$ and $4\times 10^{-6}$, respectively. The deviation between the numerical results, obtained with the two different time step sizes is sufficiently small to consider the results of the present work as independent of the time step size.

\subsection{Transient results}

This section addresses the transient behavior of a spherical bubble accelerating from rest in a quiescent viscoelastic fluid.
The bubble is driven by the force of buoyancy, which is counteracted by the viscous stress and components of the polymer stress tensor of the viscoelastic fluid, acting in the neighborhood of the fluid interface as a resistance to the bubble motion. Directly at the fluid interface, the stress is balanced by the surface tension force, according to the interface jump condition (\ref{jmomentumeqVV03}). Hence, if the stress components in interface normal direction become sufficiently large compared to the surface tension force, the bubble may deform and become non-spherical. Every change in the bubble shape in turn affects the local stress distribution around the fluid interface. The non-trivial interactions between the transient deformation of the bubble and the transient evolution of the local stress distribution around the fluid interface result in a non-monotonous behavior of the transient rise velocity. Depending on the material properties of the viscoelastic fluid, the transient rise velocity response curve may run through several local minima and maxima until a steady state is reached. 
Furthermore, the timescale of these dynamics before the transition towards the terminal rise velocity is relatively long. Regarding the velocity jump discontinuity, a question of considerable importance is whether the dynamics in the rise velocity and the transient deformation-stress interactions at the fluid interface show a substantial difference for the subcritical and the supercritical bubble volume. To obtain a more precise picture, the transient rise velocity is studied for several subcritical and supercritical bubble volumes. The bubble shape is evaluated at some characteristic points in the bubble rise velocity graph.

\begin{figure}[htbp!]
\centering{
\begin{tabular}{@{}c@{}}
{\includegraphics[width=212pt]{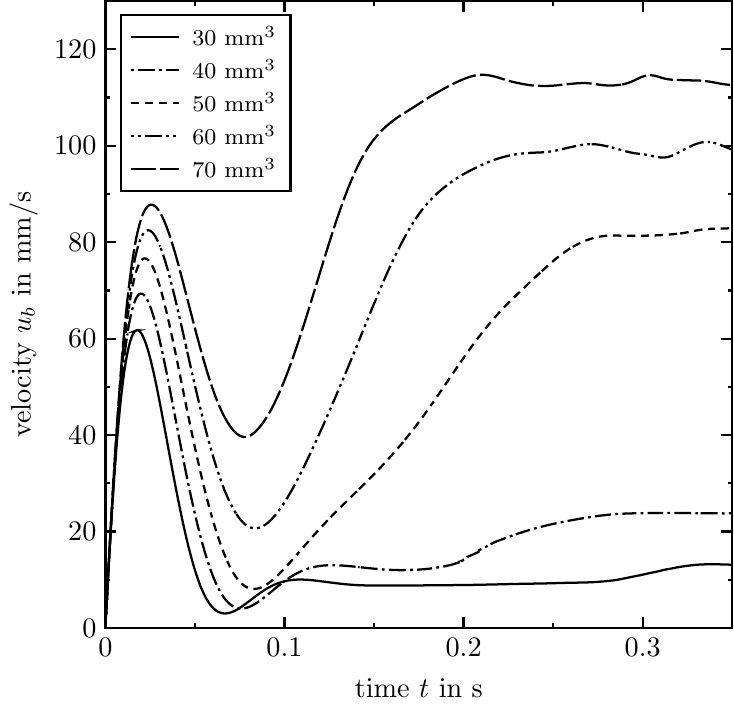}}
\end{tabular}
}
\caption{\small Transient rise velocity $u_b$ of bubbles rising in a viscoelastic fluid. The bubble volume is varied from $30$ to $70$ $\textnormal{mm}^3$. Between $40$ $\textnormal{mm}^3$ and $50$ $\textnormal{mm}^3$, the terminal rise velocity increases by a factor of $\approx 3.5$. }
\label{fig:transientVelocity}
\end{figure}

Figure \ref{fig:transientVelocity} shows the transient rise velocities as functions of time.

Results for five different bubble volumes are plotted, from which the two smaller volumes are subcritical while the three larger volumes are supercritical. This becomes obvious from the large difference in the terminal bubble rise velocities between the volumes $40$ $\textnormal{mm}^3$ and $50$ $\textnormal{mm}^3$. The comparison shows that the initial rise behavior is dominated by buoyancy; all bubbles accelerate from rest which results in a monotonous increase of the rise velocity. The slope of the rise velocity curve becomes steeper as the volume is increased. With progressing time, the acceleration decreases in size and even becomes negative. The rise velocity runs through a local maximum and decreases in size. This decrease of the rise velocity indicates that the stress layers of the viscoelastic fluid around the bubble become large enough to influence the motion of the bubble. The increasingly pronounced elastic stress response of the fluid changes the local flow field in the neighborhood of the bubble, which results in this overall deceleration of the bubble. The results suggest that this period is also crucial for the jump discontinuity in the bubble rise velocity. The differences between the subcritical and the supercritical states start to become more and more obvious. For the subcritical volumes, the rise velocity decreases to smaller values and, after passing a local minimum, remains on a relatively small level. For the supercritical volumes, the local minimum in the rise velocity is reached at higher absolute values and is followed by a second period of strong acceleration, giving rise to a steep increase in the transient rise velocity. This difference eventually results in the velocity jump discontinuity, as the rise velocity asymptotically approaches a significantly higher terminal value.

\begin{figure}[htbp!]
\centering{
\begin{tabular}{@{}c@{}}
{\includegraphics[width=172pt]{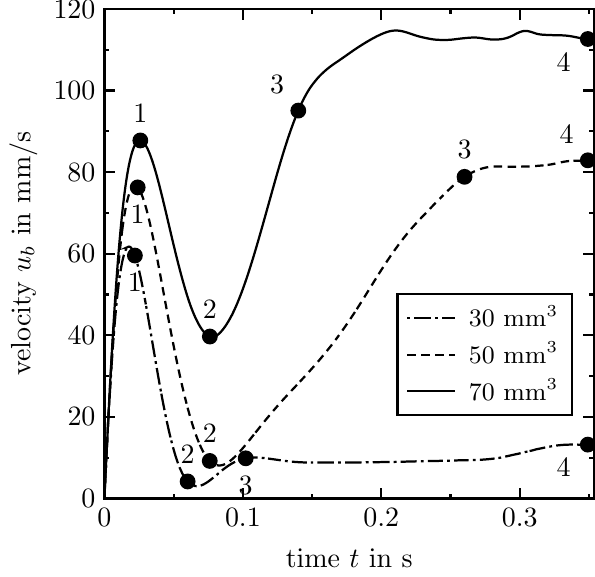}}\\
\begin{tabular}{@{}lcccc@{}}
&
\small 1 &
\small 2 &
\small 3 &
\small 4 \\
\makebox[30pt][l]{\hspace*{0cm}\raisebox{40pt}{$30$ $\textnormal{mm}^3$}} &
\includegraphics[width=80pt]{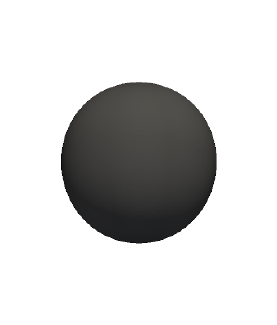} \  &
\includegraphics[width=80pt]{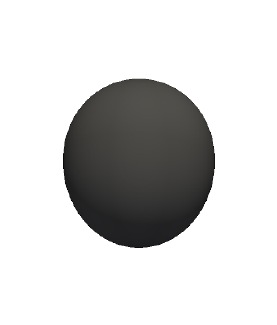} \  &
\includegraphics[width=80pt]{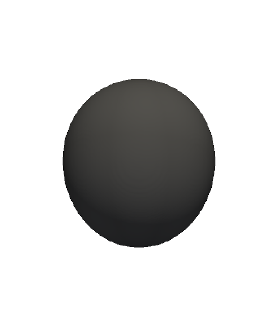} \  &
\includegraphics[width=80pt]{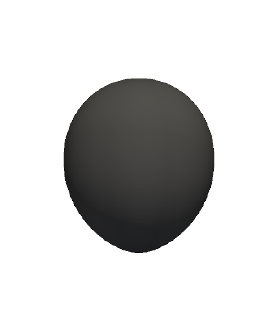} \\
\makebox[40pt][l]{\hspace*{0cm}\raisebox{40pt}{$50$ $\textnormal{mm}^3$}} &
\includegraphics[width=80pt]{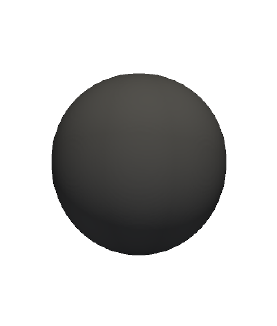} \  &
\includegraphics[width=80pt]{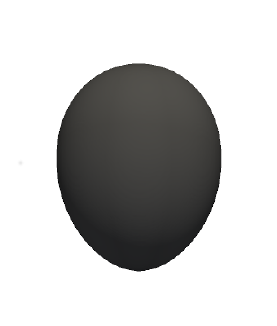} \  &
\includegraphics[width=80pt]{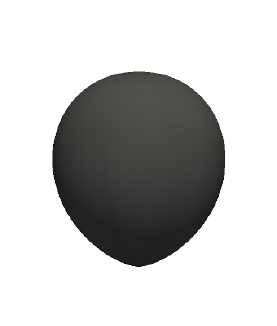} \  &
\includegraphics[width=80pt]{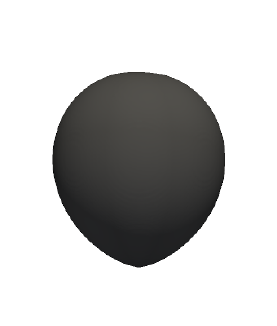} \\
\makebox[40pt][l]{\hspace*{0cm}\raisebox{40pt}{$70$ $\textnormal{mm}^3$}} &
\includegraphics[width=80pt]{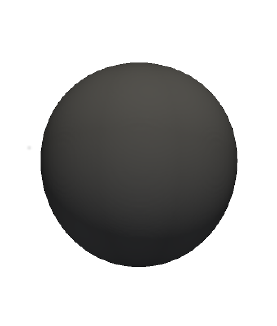} \  &
\includegraphics[width=80pt]{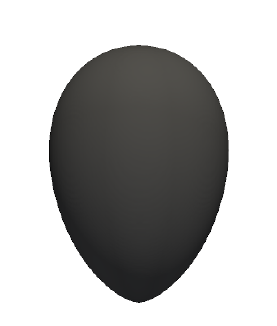} \  &
\includegraphics[width=80pt]{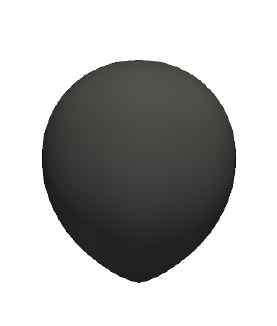} \  &
\includegraphics[width=80pt]{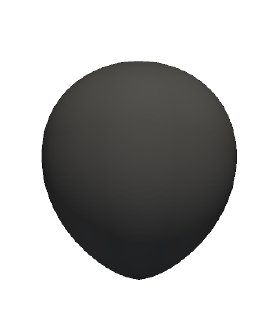}
\end{tabular}
\end{tabular}
}
\caption{\small Transient evolution of the bubble shape for three different volumes rising in a viscoelastic fluid. The small volume $30$ $\textnormal{mm}^3$ is subcritical, while the two larger volumes $50$ $\textnormal{mm}^3$ and $70$ $\textnormal{mm}^3$ are supercritical. For each volume, the respective bubble shapes are visualized at four characteristic instants in time. Note that the interface is represented as a constant iso-surface of the volume fraction field and therefore does not correspond to the exact position of the interface, which cannot be determined precisely in the algebraic VoF method.}
\label{fig:transientShape}
\end{figure}

Figure \ref{fig:transientShape} gives an impression of the dependence between the transient rise velocity and the bubble shape for different bubble volumes.

The interface is represented as the $0.5$ isosurface of the volume fraction field. Three different bubble volumes are visualized at four characteristic points in time. An overall comparison reveals that the transient change of the bubble shape becomes more pronounced as the volume is increased. In particular, the subcritical volume with $30$ $\textnormal{mm}^3$ experiences relatively small changes of the bubble shape, compared to the two supercritical volumes, $50$ $\textnormal{mm}^3$ and $70$ $\textnormal{mm}^3$. The transient rise behavior can be divided into different zones with respect to the deformation of the bubble. At the beginning of the simulation, each bubble is initialized as a sphere. As the bubble accelerates, the changes of the bubble shape are relatively insignificant. The bubble becomes slightly ellipsoidal before reaching the first local maximum in the rise velocity (point $1$ in the diagram). From this point on, the changes in the bubble shape become increasingly rapid. This increased deformation of the bubble seems to be closely related to the formation of a remarkable stress layer around the bubble interface. The bubble starts to be stretched in the vertical direction, while the rise velocity decreases. The maximum vertical length of the bubble is reached roughly at point 2. This point also represents a local minimum in the bubble rise velocity. Between point 2 and point 3, the bubble contracts significantly in the vertical direction, while the rise velocity increases again. For the supercritical bubbles, the particular shape of the trailing end appears increasingly pronounced between points 2 and 3. As time elapses further, the overall dimensions of the bubble do not change considerably. The bubble rise velocity slightly oscillates, at least for supercritical bubbles for which we observe a slightly asymmetric motion. The respective bubble shapes in point 4 represent the steady state results in this work.

\begin{figure}[h]
\centering{
\begin{tabular}{@{}cc@{}}
\makebox[172pt][l]{\hspace*{0cm}\raisebox{-86pt}{\includegraphics[width=172pt]{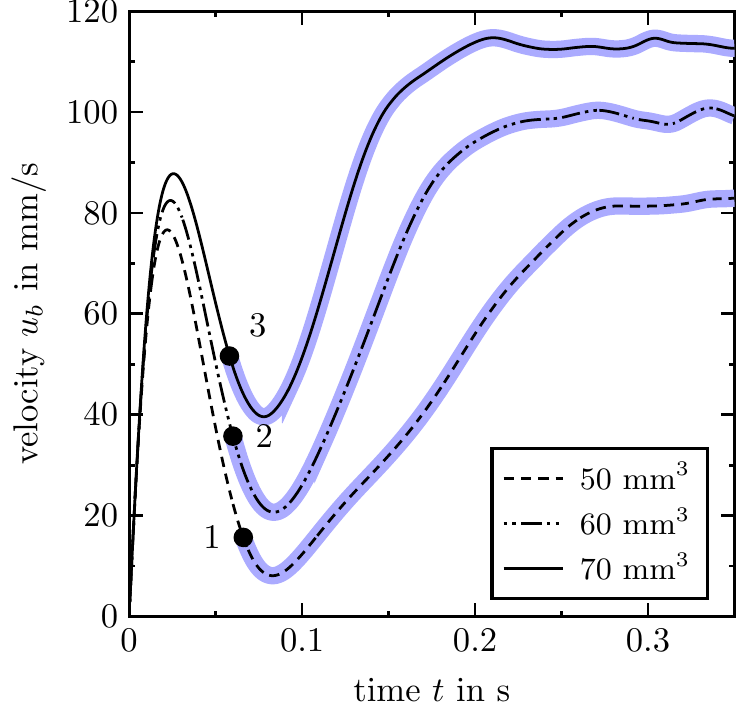}}} &
\begin{tabular}{@{}ccc@{}}
\small 1 &
\small velocity vectors \ \ &
\small radial velocity $u_r$ \\
&
\includegraphics[width=90pt]{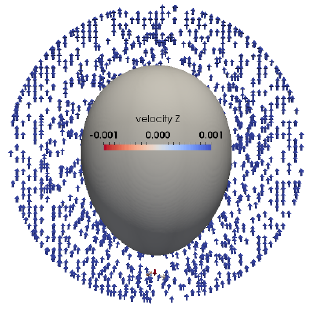} \  &
\includegraphics[width=90pt]{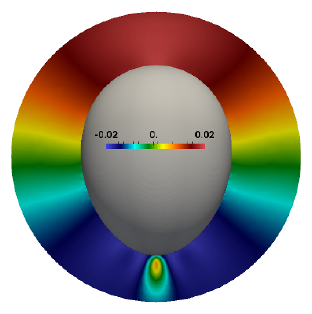} \\
\small 3 &
\small velocity vectors \ \ &
\small radial velocity $u_r$ \\
&
\includegraphics[width=90pt]{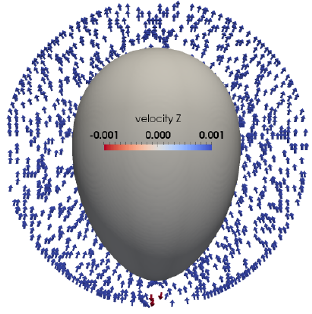} \  &
\includegraphics[width=90pt]{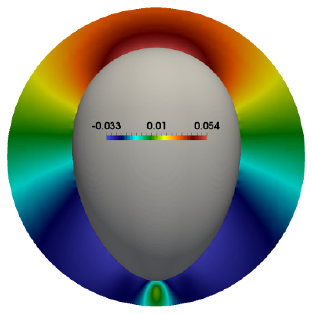}
\end{tabular}
\end{tabular}
}
\caption{\small Onset of the negative wake in the transient bubble rise velocity. Left: In the diagram the rise velocity of three supercritical bubbles is plotted over time. The presence of the negative wake is highlighted in blue. Right: Flow fields of the rising bubble at points 1 and 3 in the diagram on the l.h.s. of the figure. The reverse flow appears first in a small region closely behind the bubbles trailing end.}
\label{fig:onset}
\end{figure}

In Figure \ref{fig:onset} we study the dependence between the transient bubble rise velocity and the negative wake behind the trailing end of the rising bubble.

The evaluation shows that the formation of the negative wake behind the bubble's trailing end is a continuous dynamic process. In the diagram on the l.h.s.\ of Figure \ref{fig:onset}, the onset of the negative wake is highlighted in the curves for the transient rise velocities. The points 1 to 3 display the time when the flow reversal in the wake of the bubble first appears, for the respective three supercritical bubble volumes. The local flow fields directly at the onset of the reversal of the velocity are shown in Figure \ref{fig:onset}. It can be seen that the reversal of the velocity appears first in a relatively small region, located directly behind the bubble's trailing end. Once this reverse flow has emerged behind the bubble's trailing end this region dynamically increases in size until the conoidal shape of the negative wake is reached in the steady state. Considering these observations, it is not surprising that there is no abrupt change in the stress field around the bubble at the onset of the reverse flow. Rather, it is evident that the stress distribution behind the bubble's trailing end is the result of a continuous dynamic process, too. For the supercritical bubbles under investigation, the onset of the reverse flow takes place in a period of time, when the local stress maxima around the fluid interface increase continuously. In conjunction with this transient stress response, the bubble experiences considerable stretching in the vertical direction and the bubble rise velocity decreases.

\subsection{Steady bubble rise velocity}
\begin{figure}[h]
\centering{
\includegraphics[width=260pt]{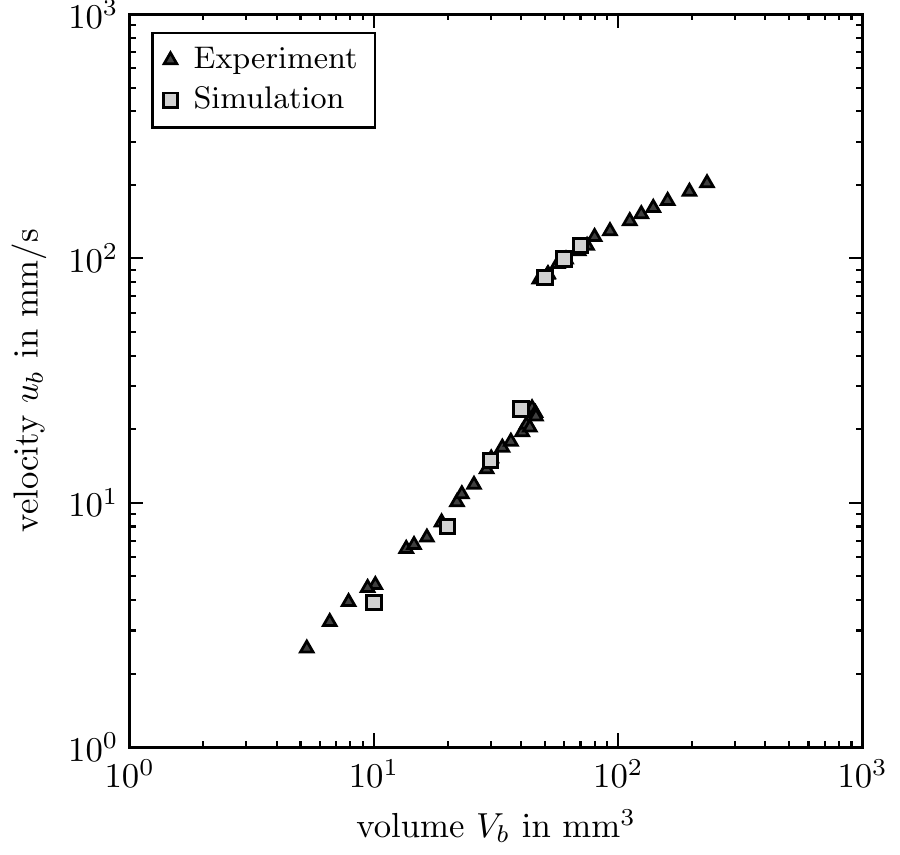}
}
\caption{\small Steady bubble rise velocity as a function of the bubble volume for bubbles rising in a viscoelastic fluid. Comparison between experiments by \cite{Pilz2007} and the numerical simulations of the present work. }
\label{fig:velvol}
\end{figure}

It has been reported that for the viscoelastic fluid under investigation, the bubbles exhibit a velocity jump discontinuity in the steady rise velocity when their volume exceeds a critical value \cite{Pilz2007}. Our goal is to predict this rather abrupt increase of the steady bubble rise velocity numerically with the proposed VoF method. A comparison of the steady rise velocity as a function of the bubble volume is shown in Figure \ref{fig:velvol}.

The simulation results obtained with the VoF method are in good quantitative agreement with the experimental measurements by Pilz and Brenn \cite{Pilz2007}. We identify a rather abrupt increase of the steady bubble rise velocity somewhere in the range between $40 \ \textnormal{mm}^3 < V_b < 50 \ \textnormal{mm}^3$, both in the experiments and the numerical results. Hence, the critical bubble volume is quantitatively predicted correctly in the numerical simulations. Also, the magnitude of the jump rise velocity is in good agreement with the experimental data: the numerical results for the steady rise velocities of the supercritical bubbles coincide with the measured values. For subcritical bubbles the simulation results show a slightly steeper increase in the rise velocity as the bubble volume is increased, compared to the experimental values.

\subsection{Negative wake}
The existence of a negative wake behind bubbles rising in viscoelastic liquids was proven by flow measurements \cite{Funfschilling2001}. The negative wake appears in the flow field which is seen by a fixed observer in the inertial reference frame. In a small distance from the bubble's trailing end, the flow direction reverses with respect to the direction of the bubble motion. This flow in reverse direction develops in a nearly conoidal region in the wake behind the bubble. Up to now, the relation between the negative wake and the velocity jump discontinuity has not been completely clarified. It is usually observed for supercritical bubbles, while very small bubbles with volumes $V_b \ll V_{b,\textnormal{crit}}$ do not show this effect. However, this does not mean that there could not exist a transition zone for subcritical bubbles, which are not significantly smaller than $V_{b,\textnormal{crit}}$, where the negative wake appears (at least temporarily), while the rise velocity remains subcritical.

\begin{figure}[htbp!]
\centering{
\begin{tabular}{@{}ccc@{}}
\small subcritical $V_b = 30$ $\textnormal{mm}^3$ &
\small supercritical $V_b = 50$ $\textnormal{mm}^3$ &
\small supercritical $V_b = 70$ $\textnormal{mm}^3$ \\
\includegraphics[width=140pt]{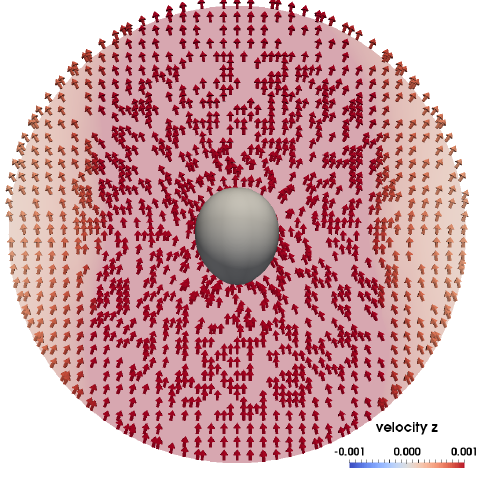} &
\includegraphics[width=140pt]{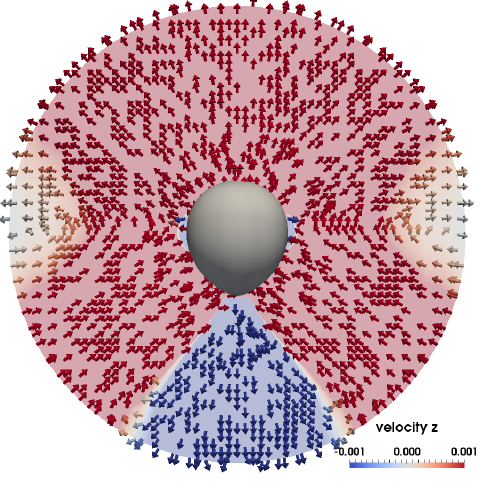} &
\includegraphics[width=140pt]{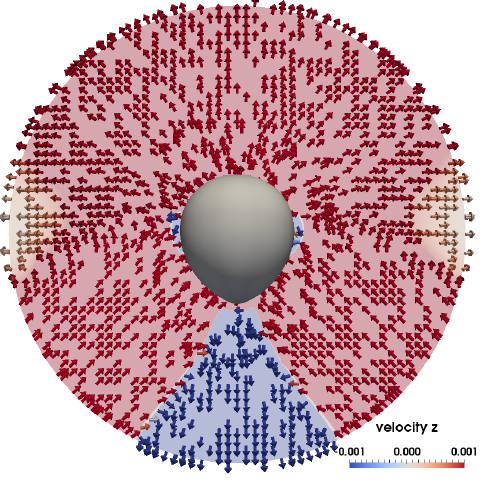}
\end{tabular}
}
\caption{\small Velocity vector field in a veritcal cutting plane through the rising bubble. The color scale is related to the vertical velocity component (in z-coordinate), where blue represents a negative vertical velocity and red represents a positive velocity in vertical direction. The negative wake behind the bubble is visible for the two supercritical states.}
\label{fig:negwake}
\end{figure}

The negative wake is also observed in the present numerical investigation of a rising bubble in a viscoelastic fluid. Figure \ref{fig:negwake} shows the velocity vector field in a vertical cutting plane through the center of the rising bubble for three different bubble volumes, as the terminal rise velocity is reached.

The smallest bubble on the l.h.s.\ of Figure \ref{fig:negwake} with $V_b = 30$ $\textnormal{mm}^3$ is subcritical. The velocity vectors in the neighborhood of the fluid interface correspond to the direction of motion of the rising bubble; they are all directed upwards. In the middle of the figure, the flow field of a supercritical bubble with $V_b = 50$ $\textnormal{mm}^3$ is visualized. In the wake behind the bubble's trailing end, the velocity vectors are oriented in reverse direction to the bubble rise velocity. The spatial dimension of the negative wake can also be seen from the color scheme, where blue represents negative velocities with respect to the vertical motion of the rising bubble. This bubble size is relatively close above the critical bubble volume of $46 \ \textnormal{mm}^3$. A larger supercritical bubble with $V_b = 70$ $\textnormal{mm}^3$ is shown on the r.h.s.\ of the figure. The velocity vectors are oriented quite similar to the smaller supercritical bubble, resulting in a qualitatively similar flow pattern. The negative wake is visible in an approximately conoidal region behind the bubble's trailing end. However, slight deviations from axial symmetry are found in the velocity field, at least for supercritical bubble volumes. Its spatial extent differs only slightly from the smaller subcritical bubble. The comparison of the flow patterns in Figure \ref{fig:negwake} suggests that, above the supercritical volume, the flow field around the rising bubble is subject to minor changes as the volume is further increased. For larger bubble volumes $V_b \gg V_{b,\textnormal{crit}}$, we expect little difference in the flow patterns. We therefore focus our investigations on the range $30 \ \textnormal{mm}^3 \leq V_b \leq 70 \ \textnormal{mm}^3$.

\begin{figure}[htbp!]
\centering{
\begin{tabular}{@{}cc@{}}
\includegraphics[width=320pt]{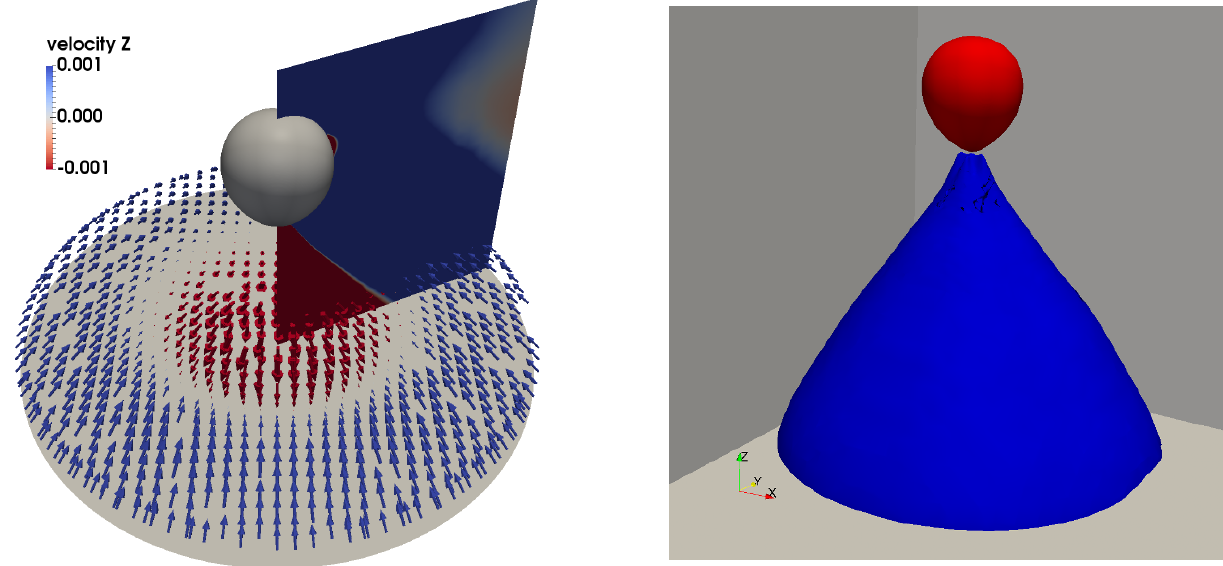}
\end{tabular}
}
\caption{\small Three-dimensional flow field of the negative wake of a supercritical bubble. Left: Velocity vectors in a horizontal plane through the bubble's wake. Right: Contour of the bubble (red) and the region of negative vertical flow (blue) behind the trailing end of the bubble.}
\label{fig:negwake3d}
\end{figure}

Figure \ref{fig:negwake3d} gives an impression of the three-dimensional extension of the negative wake behind the bubble's trailing end.

On the l.h.s.\ the velocity vectors are plotted in a horizontal cutting plane through the wake behind the bubble. In a circular region around the center of the plane, the velocity vectors point downwards, representing the negative wake. In a lateral surface around this region, the velocity is positive, representing the upwards directed flow that pushes the bubble in the vertical direction. On the r.h.s.,\ the three-dimensional contour of the negative velocities is plotted for the supercritical bubble with $V_b = 70$ $\textnormal{mm}^3$. Although the flow field slightly deviates from axial symmetry, this visualization shows the nearly conoidal shape of the negative wake.

\begin{figure}[htbp!]
\centering{
\begin{tabular}{@{}ccc@{}}
$u_r$ &
$u_\theta$ &
\small  \\
\includegraphics[width=124pt]{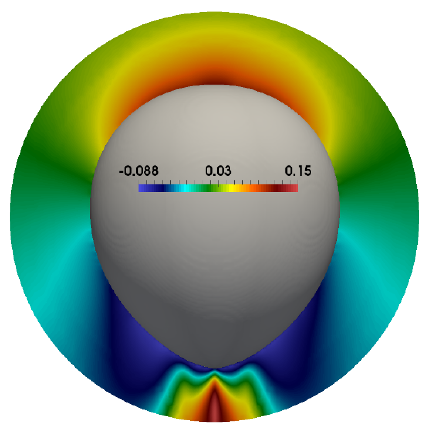} &
\includegraphics[width=124pt]{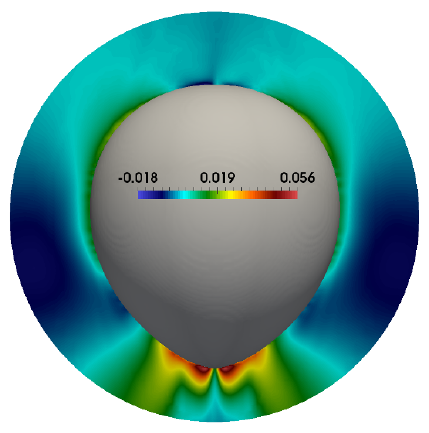} &
\includegraphics[width=140pt]{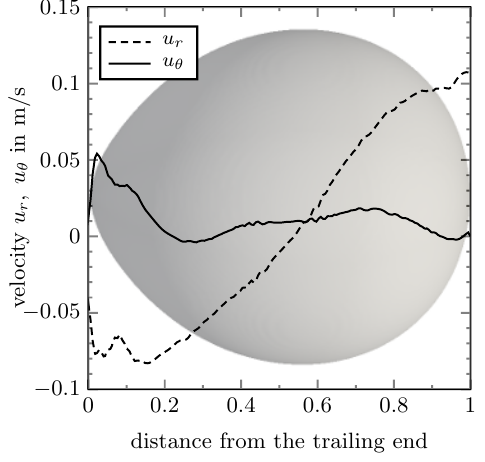}
\end{tabular}
}
\caption{\small Distribution of the velocity components $u_r$ and $u_\theta$ in spherical coordinates around a supercritical rising bubble. Left: Visualization of the velocity components in a vertical plane through the bubble center. Right: Diagram of the local velocity distribution along the cutting line between the vertical plane and the fluid interface. Note that the interface is represented as an iso-line of the volume fraction field and therefore does not correspond to the exact position of the interface, which cannot be determined precisely in the algebraic VoF method. The diagram provides a rather qualitative impression of the local velocity distribution around the bubble. }
\label{fig:negwakeVelocity}
\end{figure}

A more quantitative representation of the velocity field around the supercritical bubble with $V_b = 70$ $\textnormal{mm}^3$ is shown in Figure \ref{fig:negwakeVelocity}. 

The velocity vector field is transformed to spherical coordinates to simplify the understanding of the distribution around the bubble. The origin of the spherical coordinate system is placed in the bubble's center of mass. On the l.h.s.\ the velocity components $u_r$ and $u_\theta$ in spherical coordinates are plotted in a vertical cutting plane through the bubble center. From these representations it can be seen that, close to the bubble's trailing end, the velocity has a local minimum in the magnitude of both components. This indicates that there could be a stagnation point, located either at the bubble's trailing end or closely behind the trailing end. However, this point cannot be resolved precisely with the algebraic VoF method used in the present work. On the r.h.s.\ of Figure \ref{fig:negwakeVelocity}, the local velocity distribution is plotted along the isoline of the volume fraction value $0.7$ in the vertical cutting plane through the bubble. We use the interface representation of an isoline, since the exact position of the interface cannot be determined accurately with an algebraic VoF method. The volume fraction value $0.7$ is chosen, because we intend to evaluate the liquid-sided limit of the viscoelastic fluid, which is represented by the volume faction value $1$. Therefore, we choose a volume fraction value between $0.5$ and $1$. It is evident that this representation allows only for a qualitative impression on the local velocity distribution around the bubble. Keeping this in mind, we observe that the local minimum in the magnitude of both velocity components is visible at position 0. As we move slightly upwards along the interface, the magnitude of both components increases steeply. This is the region where a strong flow is pushing the bubble upwards. Moving further towards the equator, the flow becomes weaker. The radial velocity component changes its sign approximately at the equator, where the horizontal bubble diameter has its maximum. In the upper hemisphere of the bubble, the flow in radial direction increases in size as we move further towards the top, while $u_\theta$ remains on a small level.

\subsection{Stress distribution}

For the investigation of the stress profiles around the fluid interface of the bubble rising in a viscoelastic fluid, we transform the stress tensor into spherical coordinates. The origin of the spherical coordinate system is placed in the bubble's center of mass. In the neighborhood of the fluid interface, the stress tensor in spherical coordinates has three important components: the two normal stress components $\tau_{r r}$ and $\tau_{\theta \theta}$ and the shear stress $\tau_{r \theta}$. These stress components are evaluated below for a subcritical volume and a supercritical volume.

\begin{figure}[htbp!]
\centering{
\begin{tabular}{@{}ccc@{}}
\multicolumn{3}{l}{(a) subcritical} \\
$\tau_{r r}$ &
$\tau_{\theta \theta}$ &
\small  \\
\includegraphics[width=124pt]{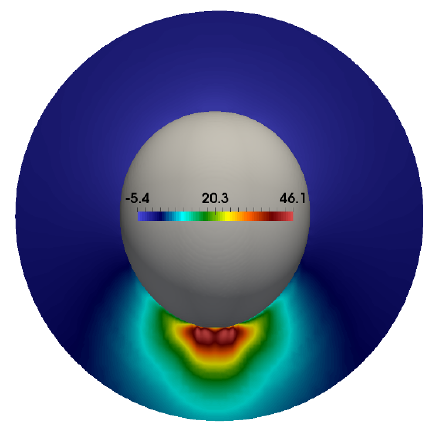} &
\includegraphics[width=124pt]{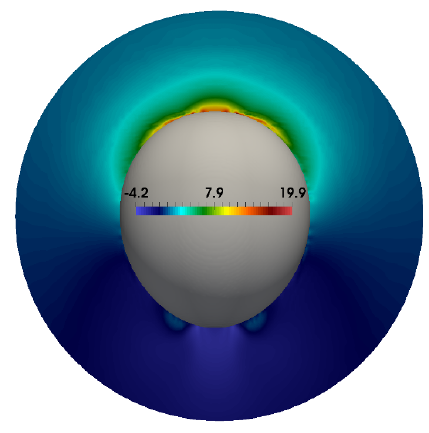} &
\includegraphics[width=140pt]{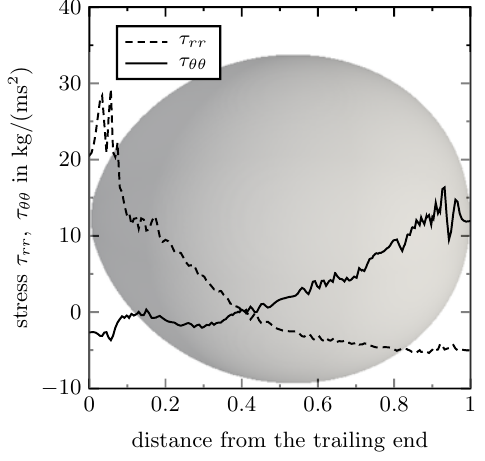} \\
\multicolumn{3}{l}{(b) supercritical} \\
$\tau_{r r}$ &
$\tau_{\theta \theta}$ &
\small  \\
\includegraphics[width=124pt]{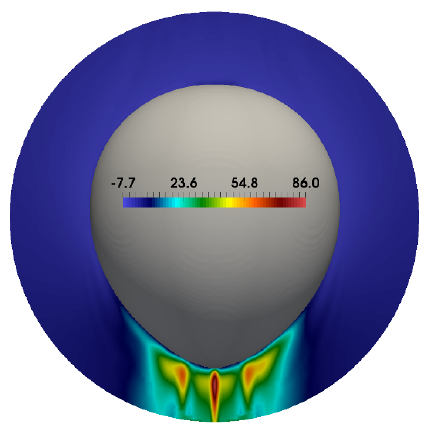} &
\includegraphics[width=124pt]{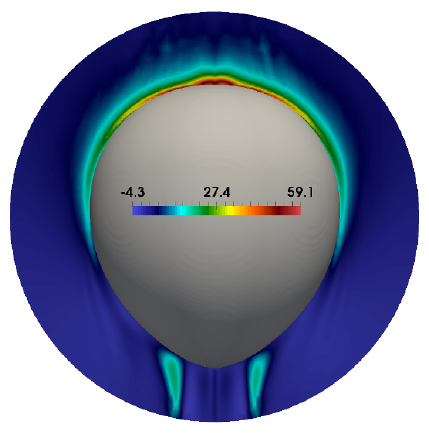} &
\includegraphics[width=140pt]{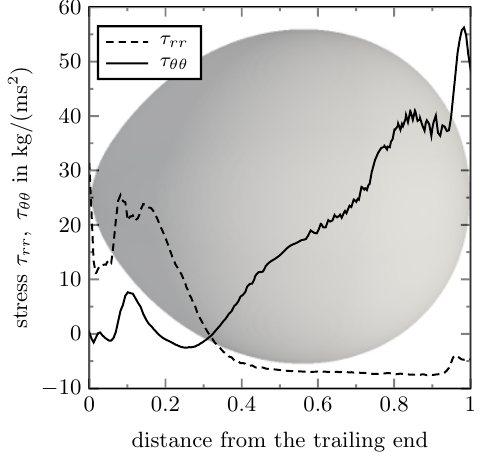}
\end{tabular}
}
\caption{\small Distribution of the normal stress components $\tau_{r r}$ and $\tau_{\theta \theta}$ in spherical coordinates around the rising bubble in a viscoelastic fluid. Comparison between a subcritcal and a supercritical state. Left: Visualization of the stress components in a vertical plane through the bubble center. Right: Diagrams of the local normal stress distribution along the cutting line between the vertical plane and the fluid interface. Note that the interface is represented as an iso-line of the volume fraction field and therefore does not correspond to the exact position of the interface, which cannot be determined precisely in the algebraic VoF method. The diagram provides a rather qualitative impression of the local normal stress distribution around the bubble. }
\label{fig:stressNormal}
\end{figure}

Figure \ref{fig:stressNormal} shows the normal stress profiles for $\tau_{r r}$ and $\tau_{\theta \theta}$ in spherical coordinates around two rising bubbles in a viscoelastic fluid.

The two bubbles have different volumes; one bubble has a subcritical volume while the other bubble volume is in the supercritical regime. On the l.h.s.,\ the stress distribution is shown in a vertical cutting plane through the bubble center. A comparison of the subcritical and supercritical bubbles shows that the normal stress in the neighborhood of the fluid interface is distributed very different by these two states. Also, the scales differ considerably. The radial stress component has a maximum behind the trailing end of the bubble. The absolute value of this maximum is significantly higher for the supercritical volume. The normal stress component $\tau_{\theta \theta}$ has a maximum in the upper hemisphere of the bubble. In this component, the difference of the maximum values for the subcritical and the supercritical bubble is even more pronounced. It is also remarkable that the spatial extent of the large positive values in $\tau_{\theta \theta}$ around the bubble interface is significantly greater for the supercritical bubble; all values in the upper bubble hemisphere are larger than the $\tau_{\theta \theta}$ maximum of the subcritical bubble. On the r.h.s.\ of Figure \ref{fig:stressNormal}, the local normal stress distribution is plotted along the isoline of the volume fraction value $0.7$ in the vertical cutting plane through the bubble. For the reasons given above, this representation allows only for a qualitative impression on the local stress distribution around the bubble. The characteristic difference between the subcritical and the supercritical state is obvious from the comparison of the $\tau_{\theta \theta}$ lines. In the upper half of the line (for distances $>$ $0.5$ from the trailing end), $\tau_{\theta \theta}$ has much greater values.

\begin{figure}[h]
\centering{
\begin{tabular}{@{}cc@{}}
\multicolumn{2}{c}{isovalue $\tau_{\theta \theta} = 12 \ \textnormal{kg}/(\textnormal{ms}^2)$ in red color} \\
subcritical  \ \ &
supercritical \\
\includegraphics[width=124pt]{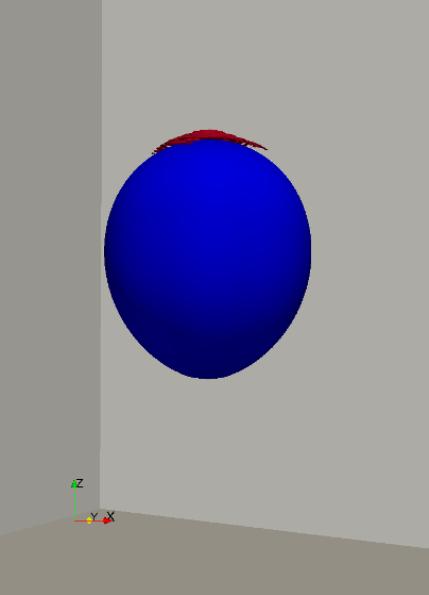} \ & \
\includegraphics[width=124pt]{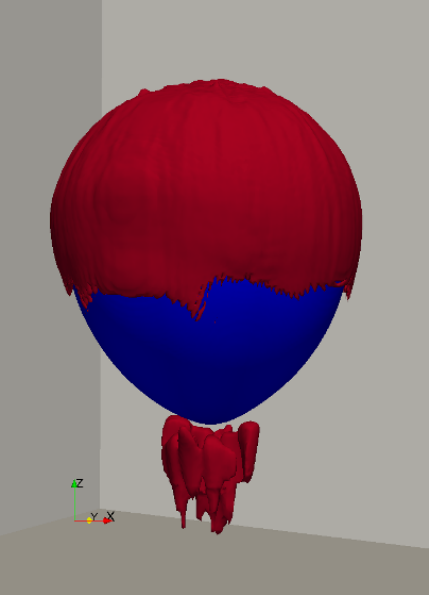}
\end{tabular}
}
\caption{\small Three-dimensional visualization of the normal stress component $\tau_{\theta \theta}$ around the bubble. The isovalue $\tau_{\theta \theta} = 12 \ \ \textnormal{kg}/(\textnormal{ms}^2)$ is used to illustrate the significant difference in the distribution around the subcritical and the supercritical bubble, respectively.}
\label{fig:stressNormalC}
\end{figure}

The difference in the distribution of the normal stress component $\tau_{\theta \theta}$ around the bubble becomes more obvious in the three-dimensional visualization in Figure \ref{fig:stressNormalC}.

This figure shows a plot of the isovalue $\tau_{\theta \theta} = 12 \ \textnormal{kg}/(\textnormal{ms}^2)$, which is slightly smaller than the maximum of the subcritical volume. In this representation, the upper hemisphere of the supercritical bubble is completely covered by a normal stress layer with large positive $\tau_{\theta \theta}$ values, while in the lower bubble hemisphere the values of $\tau_{\theta \theta}$ are smaller than $12 \ \textnormal{kg}/(\textnormal{ms}^2)$. Another local maximum in the $\tau_{\theta \theta}$ component is located in the wake of the supercritical bubble. From a comparison with the cutting plane in Figure \ref{fig:stressNormal} it becomes obvious that this local maximum forms approximately a torus around the region of the negative wake behind the trailing end of the bubble.

\begin{figure}[htbp!]
\centering{
\begin{tabular}{@{}ccc@{}}
\multicolumn{1}{c}{subcritical} & \multicolumn{1}{c}{supercritical}\\
$\tau_{r \theta}$ &
$\tau_{r \theta}$ &
\small  \\
\includegraphics[width=124pt]{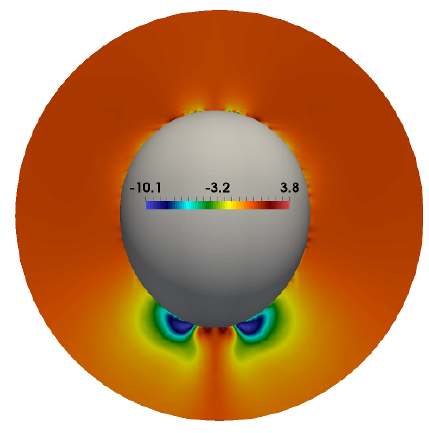} &
\includegraphics[width=124pt]{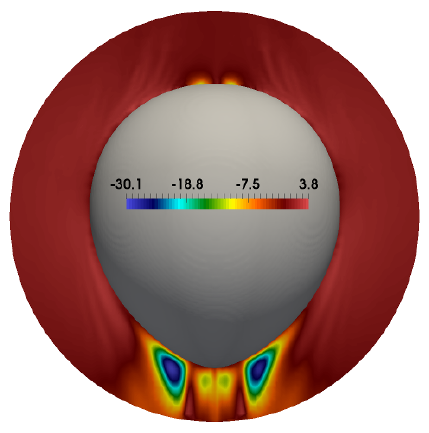} &
\includegraphics[width=140pt]{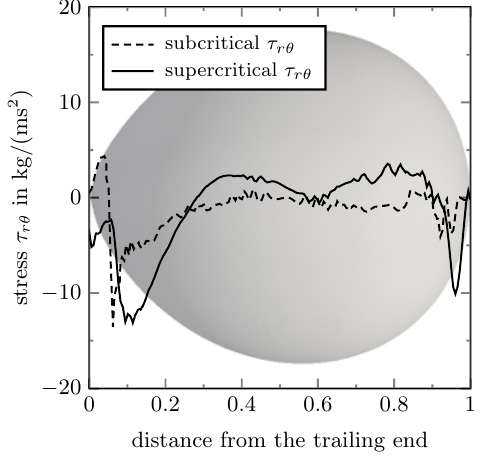}
\end{tabular}
}
\caption{\small Distribution of the shear stress component $\tau_{r \theta}$ in spherical coordinates around the rising bubble in a viscoelastic fluid. Left: Comparison between a subcritcal and a supercritical state. Right: Diagram of the local shear stress distribution along the cutting line between along the cutting line between the vertical plane and the fluid interface. Note that the interface is represented as an iso-line of the volume fraction field and therefore does not correspond to the exact position of the interface, which cannot be determined precisely in the algebraic VoF method. The diagram provides a rather qualitative impression of the local shear stress distribution around the bubble.}
\label{fig:stressShear}
\end{figure}

Figure \ref{fig:stressShear} shows the shear stress component $\tau_{r \theta}$ in spherical coordinates.

In the upper hemisphere of the bubble, the magnitude of the shear stress in the neighborhood of the fluid interface is rather small for both, the subcritical and the supercritical volume. Comparing with the normal stress $\tau_{\theta \theta}$, we can state that the upper hemisphere of the bubble is dominated by the component $\tau_{\theta \theta}$, while the shear stress is insignificant. In the lower hemisphere of the bubble, the magnitude of the shear stress $\tau_{r \theta}$ remains relatively small over a wide range along the interface. Only in the lower quarter of the hemisphere, a considerable amount of shearing is present. An extremum is located in a relatively small region of the upwards directed flow zone, close around the outer dimension of the negative wake. In this region, $\tau_{r \theta}$ has negative values.

The extra stress components displayed in Figs.\ \ref{fig:stressNormal} to \ref{fig:stressShear} are due to the interplay between the solvent flow field and the deformation of the polymer molecules in the flow around the bubble. Following the fluid approaching the upstream stagnation point of the bubble, we see the polymer molecules stretched in the $\theta$-direction, according to the biaxial straining flow around the upper region of the bubble, leading to the dominant normal stress component $\tau_{\theta \theta}$ \cite{Noh1993}. In this flow, the polymer molecules are stretched into a deformed state, from which they tend to relax on the time scale of the stress relaxation time. In the lower region of the bubble, downstream from the rear bubble end, we see a uniaxial state of straining flow. In that flow, the polymers are stretched again, this time in the radial direction. The related normal stress deforms the bubble into its elongated form, seen both in the simulation and in the experiment. In comparison to the normal stress components, the shear stress $\tau_{r \theta}$ remains small throughout the flow field near the bubble surface. Considering the shear stress continuity across the gas-liquid interface, this is not surprising, given the dynamic viscosity of the air which is four orders of magnitude smaller than the one of the liquid, and the shear rate in the bubble of the order of $2 \ \textnormal{s}^{-1}$ at a bubble velocity and size of $10 \ \textnormal{mm}/\textnormal{s}$ and $5 \ \textnormal{mm}$, respectively. Furthermore, the spatial distribution of the stresses along the bubble surface, in particular their persistence after build-up due to changes in the flow field, depends on the Deborah number $\tau u_b / d_b$. Larger Deborah numbers tend to conserve the stresses farther downstream from the upper stagnation point.

\section{Summary and Conclusions}
An extended VoF method was proposed for the three-dimensional DNS of two-phase flows of a Newtonian and a viscoelastic fluid. The VoF method is based on a one-field formulation of the volume-averaged two-phase flow equations. Special care was taken on the derivation of a closure for the polymer stress tensor for this one-field formulation. To improve the stability at moderate and high Weissenberg numbers, a change-of-variable representation for the constitutive equations of the polymer stress was incorporated. This general representation includes different rheological models, such as the Oldroyd-B and the PTT models.

Numerical experiments of a rising bubble in a viscoelastic fluid were carried out to demonstrate the capability of the flow solver to predict characteristic flow phenomena, such as the rise velocity jump discontinuity and the negative wake. It is particularly remarkable that good quantitative agreement between the numerical results and experimental measurements was achieved. In this numerical study, the velocity jump discontinuity appears at a similar critical volume and has the same magnitude as in the experimental reference data. Furthermore, the negative wake was shown in the flow patterns of the supercritical rising bubbles, which qualitatively corresponds to the experimental observations reported in literature.

The numerical results for the rising bubble in a viscoelastic fluid provide detailed insight into the local flow patterns and the stress distribution in the neighborhood of the rising bubble. Slight deviations from axial symmetry in the velocity and stress fields for supercritical bubble volumes substantiate the application of a fully three-dimensional approach. The results suggest that the distribution of the normal stress tensor component $\tau_{\theta\theta}$ around the fluid interface is significantly different for subcritical and supercritical bubble volumes. For supercritical volumes, the upper hemisphere of the bubble is completely covered by a stress layer of relatively large positive $\tau_{\theta\theta}$ values, while such a stress layer is not observed for subcritical bubbles. We interpret this normal stress layer as a zone in which the polymer molecules are being stretched in the tangential direction along the fluid interface. If we consider a relative coordinate system aligned with the motion of the rising bubble, the polymer molecules move around the bubble from the top to the bottom. In the upper hemisphere, the polymer is stretched as it moves along the fluid interface. This stored energy is released on the lower hemisphere, where $\tau_{\theta\theta}$ is very small in magnitude. Potentially, this contraction of the polymer molecules in the lower hemisphere of the bubble contributes to the rise velocity. Further evidence of this will be provided in a forthcoming study.

\section*{Acknowledgements}
This work was partly supported from the German Research Foundation (DFG) within the collaborative research center SFB 1194 ``Interaction between Transport and Wetting Processes". This support is gratefully acknowledged.
Extensive calculations for this research were conducted on the Lichtenberg high performance computer of the Technische Universit{\"a}t Darmstadt.

\bibliographystyle{amsplain}
\bibliography{LIT/bibliography}

\appendix
\section{Computation of the tensors $\BT$ and $\OmegaT$}
\label{asec:lddt}
Using the diagonalization $\CT = \QT \dprod \LambdaT \dprod \trans{\QT}$, the tensor $\BT$ can be computed as $\BT = \QT \dprod \tilde{\BT} \dprod \trans{\QT}$, where the elements of the diagonal tensor $\tilde{\BT}$ are given as a function of the tensor $\LT = \QT \dprod \tilde{\LT} \dprod \trans{\QT}$ as $\tilde{b}_{ii} = \tilde{l}_{ii}$. The tensor $\OmegaT$ can be computed as $\OmegaT = \QT \dprod \tilde{\OmegaT} \dprod \trans{\QT}$, where the tensor $\tilde{\OmegaT}$ has zero diagonal entries $\tilde{\omega}_{ii} = 0$, while its off-diagonal elements are given by 
\begin{equation}
\label{diagonalizeCr82}
\tilde{\omega}_{{{ij}, \; {i \neq j}}} = \frac{\lambda_{ii} {\tilde{l}_{{ij}, \; {i \neq j}}} + \lambda_{jj} {\tilde{l}_{{ji}, \; {j \neq i}}}}{\lambda_{jj} - \lambda_{ii}}, \ i, j = 1, 2, 3.
\end{equation}
For a detailed description on the local decomposition of the deformation terms in the convective derivative, we refer to \cite{Niethammer2017}.

\section{Mixture conformation tensor equation}
\label{asec:mcte}
Summing up the volume averaged conformation tensor equations of phase $\phaseOne$ and phase $\phaseTwo$, we find
\begin{align}
\nonumber
&\partial_t \alpha_\phaseOne \overline{\CT}^\phaseOne + 
\partial_t \alpha_\phaseTwo \overline{\CT}^\phaseTwo +
\left( \overline{\velocity}^\phaseOne \dprod \nabla \right) \alpha_\phaseOne \overline{\CT}^\phaseOne 
+ \left( \overline{\velocity}^\phaseTwo \dprod \nabla \right) \alpha_\phaseTwo \overline{\CT}^\phaseTwo \\
\nonumber
&- \alpha_\phaseOne \overline{\LT}^\phaseOne \dprod \overline{\CT}^\phaseOne
- \alpha_\phaseTwo \overline{\LT}^\phaseTwo \dprod \overline{\CT}^\phaseTwo
- \alpha_\phaseOne \overline{\CT}^\phaseOne \dprod \trans{(\overline{\LT}^\phaseOne)} 
- \alpha_\phaseTwo \overline{\CT}^\phaseTwo \dprod \trans{(\overline{\LT}^\phaseTwo)} \\
\label{aconstCEqva01}
&=
 {\frac{\alpha_\phaseOne}{\tau_\phaseOne} \cvec{P}\left(\overline{\CT}^\phaseOne \right)}
+ {\frac{\alpha_\phaseTwo}{\tau_\phaseTwo} \cvec{P}\left(\overline{\CT}^\phaseTwo \right)}.
\end{align}
Our goal is to express (\ref{aconstCEqva01}) in terms of the mixture quantities $\velocity_m$, $\CT_m$ and the relative velocity $\velocity_r$.
With the definitions of the mixture velocity (\ref{mixvelocity00}), the mixture conformation tensor $\CT_m = \alpha_\phaseOne \overline{\CT}^\phaseOne + \alpha_\phaseTwo \overline{\CT}^\phaseTwo$ and the relative velocity $\velocity_r = \overline{\velocity}^\phaseTwo - \overline{\velocity}^\phaseOne$, we find
\begin{align}
\nonumber
&\partial_t {\CT}_m + \left( \velocity_m \dprod \nabla \right){\CT}_m
+ \nabla \dprod \left[\velocity_r \left(\frac{\alpha_{\phaseOne} \alpha_{\phaseTwo} \rho_\phaseOne}{\rho_m} \overline{\CT}^\phaseTwo - \frac{\alpha_\phaseOne \alpha_\phaseTwo \rho_\phaseTwo}{\rho_m} \overline{\CT}^\phaseOne \right)\right] \\
\nonumber
&- {\LT}_m \dprod {\CT}_m - \LT_{\phaseOne, r} \dprod \alpha_\phaseTwo \overline{\CT}^\phaseTwo
+ \LT_{\phaseTwo, r} \dprod \alpha_\phaseOne \overline{\CT}^\phaseOne
  \\
\nonumber
&- {\CT}_m \dprod \trans{({\LT}_m)} 
- \alpha_\phaseTwo \overline{\CT}^\phaseTwo \dprod \trans{(\LT_{\phaseOne, r})}
+ \alpha_\phaseOne \overline{\CT}^\phaseOne \dprod \trans{(\LT_{\phaseTwo, r})}
\\
\label{aconstCEqva01d}
&=
 {\frac{\alpha_\phaseOne}{\tau_\phaseOne} \cvec{P}\left(\overline{\CT}^\phaseOne \right)}
+
 {\frac{\alpha_\phaseTwo}{\tau_\phaseTwo} \cvec{P}\left(\overline{\CT}^\phaseTwo \right)}
,
\end{align}
where the definitions for the gradient terms of the relative velocity read
\begin{align}
\LT_{\phaseOne, r} &:= \trans{\left[\nabla \left(\frac{\alpha_\phaseOne \rho_\phaseOne}{\rho_m} \velocity_r \right)\right]} - \frac{\zeta}{2} \left({\left[\nabla \left(\frac{\alpha_\phaseOne \rho_\phaseOne}{\rho_m} \velocity_r \right)\right]} + \trans{\left[\nabla \left(\frac{\alpha_\phaseOne \rho_\phaseOne}{\rho_m} \velocity_r \right)\right]} \right)\\
\LT_{\phaseTwo, r} &:= \trans{\left[\nabla \left(\frac{\alpha_\phaseTwo \rho_\phaseTwo}{\rho_m} \velocity_r \right)\right]} - \frac{\zeta}{2} \left({\left[\nabla \left(\frac{\alpha_\phaseTwo \rho_\phaseTwo}{\rho_m} \velocity_r \right)\right]} + \trans{\left[\nabla \left(\frac{\alpha_\phaseTwo \rho_\phaseTwo}{\rho_m} \velocity_r \right)\right]} \right).
\end{align}
With the mixture model assumption $\velocity_r = \vec{0}$ (\ref{aconstCEqva01d}) reduces to
\begin{align}
\nonumber
&\partial_t {\CT}_m + \left( \velocity_m \dprod \nabla \right){\CT}_m
- {\LT}_m \dprod {\CT}_m - {\CT}_m \dprod \trans{({\LT}_m)} 
\\
\label{aconstCEqva01e}
&=
 {\frac{\alpha_\phaseOne}{\tau_\phaseOne} \cvec{P}\left(\overline{\CT}^\phaseOne \right)}
+
 {\frac{\alpha_\phaseTwo}{\tau_\phaseTwo} \cvec{P}\left(\overline{\CT}^\phaseTwo \right)}
.
\end{align}

\end{document}